\documentclass[useAMS,usenatbib]{mn2e}

\def\aap{AA}

\def\apjl{ApJL}

\def\mnras{MNRAS}
\def\apj{ApJ}

\def\aj{AJ}
\def\jcap{JCAP}

\def\pasj{PASJ}
\def\procspie{Proc.~SPIE}
\def\nat{Nat}

\def\zs{{z_{\rm s}}}
\def\zl{{z_{\rm l}}}
\def\ase{{\prime\prime}}
\def\Jxy{{J0408-5354}}

\usepackage{graphicx}
\usepackage{float}
\usepackage{amssymb}
\usepackage{amsfonts}
\usepackage{amsmath} 
\usepackage{color}

\def\aaemail{\tt aagnello@eso.org}
\def\eso{European Southern Observatory, Karl-Schwarzschild-Strasse 2, 85748 Garching bei M{\"u}nchen, DE}
\def\ucla{Department of Physics and Astronomy, PAB, 430 Portola Plaza, Box 951547, Los Angeles, CA 90095-1547, USA}
\def\kipac{Kavli Institute for Particle Astrophysics and Cosmology, Stanford University, 452 Lomita Mall, Stanford, CA 94305, USA}
\def\asiaa{Institute of Astronomy and Astrophysics, Academia Sinica, P.O. Box 23-141, Taipei 10617, Taiwan}
\def\mpa{Max-Planck-Institut für Astrophysik, Karl-Schwarzschild-Str. 1, D-85741 Garching, Germany}
\def\ucd{Department of Physics, University of California Davis, 1 Shields Avenue, Davis, CA 95616, USA}
\def\asiaa{Institute of Astronomy and Astrophysics, Academia Sinica, P.O.~Box 23-141, Taipei 10617, Taiwan}
\def\ioa{Institute of Astronomy, Madingley Road, Cambridge CB3 0HA, UK}
\def\kavli{Kavli Institute for Cosmology, University of Cambridge, Madingley Road, Cambridge CB3 0HA, UK}
\def\braz{CAPES Foundation, Ministry of Education of Brazil, Bras{\'i}lia - DF 70040-020, Brazil}
\def\cbpf{ICRA, Centro Brasileiro de Pesquisas Físicas, Rua Dr. Xavier Sigaud 150, CEP 22290-180, Rio de Janeiro, RJ, Brazil}
\def\braza{Laborat\'orio Interinstitucional de e-Astronomia - LIneA, Rua Gal. Jos\'e Cristino 77, Rio de Janeiro, RJ - 20921-400, Brazil}
\def\brazb{Observat\'orio Nacional, Rua Gal. Jos\'e Cristino 77, Rio de Janeiro, RJ - 20921-400, Brazil}
\def\brazc{Departamento de F\'{\i}sica Matem\'atica,  Instituto de F\'{\i}sica, Universidade de S\~ao Paulo,  CP 66318, CEP 05314-970, S\~ao Paulo, SP,  Brazil}

\def\ipmu{Kavli IPMU (WPI), UTIAS, The University of Tokyo, Kashiwa, Chiba 277-8583, Japan}
\def\mit{MIT Kavli Institute for Astrophysics and Space Research, 37-664G, 77 Massachusetts Avenue, Cambridge, MA 02139}
\def\fnal{Fermi National Accelerator Laboratory, Batavia, IL 60510}
\def\epfl{Laboratoire d'Astrophysique, Ecole Polytechnique F\'ed\'erale de Lausanne (EPFL), Observatoire de Sauverny, CH-1290 Versoix, Switzerland}
\def\hyder{Department of Physics, IIT Hyderabad, Kandi, Telangana 502285, India}
\def\slac{Kavli Institute for Particle Astrophysics and Cosmology, Stanford University, 452 Lomita Mall, Stanford, CA 94035, USA}
\def\illa{Department of Astronomy, University of Illinois, 1002 W. Green Street, Urbana, IL 61801, USA}
\def\illb{National Center for Supercomputing Applications, 1205 West Clark St., Urbana, IL 61801, USA}
\def\espa{Institut de Ci\`encies de l'Espai, IEEC-CSIC, Campus UAB, Carrer de Can Magrans, s/n,  08193 Bellaterra, Barcelona, Spain}
\def\espb{Institut de F\'{\i}sica d'Altes Energies, Universitat Aut\`onoma de Barcelona, E-08193 Bellaterra, Barcelona, Spain}
\def\cnrs{CNRS, UMR 7095, Institut d'Astrophysique de Paris, F-75014, Paris, France}
\def\cata{Instituci\'o Catalana de Recerca i Estudis Avan\c{c}ats, E-08010 Barcelona, Spain}
\def\espc{Centro de Investigaciones Energ\'eticas, Medioambientales y Tecnol\'ogicas (CIEMAT), Madrid, Spain}
\def\espd{Instituto de Fisica Teorica UAM/CSIC, Universidad Autonoma de Madrid, 28049 Madrid, Spain}
\def\eth{Institute for Astronomy, Department of Physics, ETH Zurich, Wolfgang-Pauli-Strasse 27, 8093, Zurich, Switzerland}
\def\exc{Excellence Cluster Universe, Boltzmannstr.\ 2, 85748 Garching, Germany}
\def\lmu{Faculty of Physics, Ludwig-Maximilians University, Scheinerstr. 1, 81679 Munich, Germany}

\def\penns{Department of Physics and Astronomy, University of Pennsylvania, Philadelphia, PA 19104, USA}
\def\jpl{Jet Propulsion Laboratory, California Institute of Technology, 4800 Oak Grove Dr., Pasadena, CA 91109, USA}
\def\mich{Department of Physics, University of Michigan, Ann Arbor, MI 48109, USA}

\def\ohioa{Center for Cosmology and Astro-Particle Physics, The Ohio State University, Columbus, OH 43210, USA}
\def\ohiob{Department of Physics, The Ohio State University, Columbus, OH 43210, USA}
\def\aao{Australian Astronomical Observatory, North Ryde, NSW 2113, Australia}
\def\texas{George P. and Cynthia Woods Mitchell Institute for Fundamental Physics and Astronomy, and Department of Physics and Astronomy, Texas A\&M University, College Station, TX 77843,  USA}
\def\slac{SLAC National Accelerator Laboratory, Menlo Park, CA 94025, USA}
\def\sorb{Sorbonne Universit\'es, UPMC Univ Paris 06, UMR 7095, Institut d'Astrophysique de Paris, F-75014, Paris, France}
\def\sussex{Department of Physics and Astronomy, Pevensey Building, University of Sussex, Brighton, BN1 9QH, UK}
\def\stanf{Department of Physics, Stanford University, 382 Via Pueblo Mall, Stanford, CA 94305, USA}
\def\ctio{Cerro Tololo Inter-American Observatory, National Optical Astronomy Observatory, Casilla 603, La Serena, Chile}
\def\ucl{Department of Physics \& Astronomy, University College London, Gower Street, London, WC1E 6BT, UK}
\def\cnrs{CNRS, UMR 7095, Institut d'Astrophysique de Paris, F-75014, Paris, France}
\def\sorb{Sorbonne Universit\'es, UPMC Univ Paris 06, UMR 7095, Institut d'Astrophysique de Paris, F-75014, Paris, France}

\def\ports{Institute of Cosmology \& Gravitation, University of Portsmouth, Portsmouth, PO1 3FX, UK}
\def\ucsb{Department of Physics, University of California, Santa Barbara, CA 93106, USA}
\def\nick{Staples High School, Westport CT}
\def\ctio{Cerro Tololo Inter-American Observatory, National Optical Astronomy Observatory, Casilla 603, La Serena, Chile}
\def\ucl{Department of Physics \& Astronomy, University College London, Gower Street, London, WC1E 6BT, UK}
\def\rhodes{Department of Physics and Electronics, Rhodes University, PO Box 94, Grahamstown, 6140, South Africa}
\def\berk{Department of Astronomy, University of California, Berkeley,  501 Campbell Hall, Berkeley, CA 94720, USA}
\def\lbnl{Lawrence Berkeley National Laboratory, 1 Cyclotron Road, Berkeley, CA 94720, USA}
\def\wash{Astronomy Department, University of Washington, Box 351580, Seattle, WA 98195, USA}
\def\prince{Department of Astrophysical Sciences, Princeton University, Peyton Hall, Princeton, NJ 08544, USA}
\def\ifae{Institut de F\'{\i}sica d'Altes Energies (IFAE), The Barcelona Institute of Science and Technology, Campus UAB, 08193 Bellaterra (Barcelona) Spain}
\def\south{School of Physics and Astronomy, University of Southampton,  Southampton, SO17 1BJ, UK}
\def\abc{"Universidade Federal do ABC, Centro de Ci\^encias Naturais e Humanas, Av. dos Estados, 5001, Santo Andr\'e, SP, Brazil, 09210-580}
\def\oak{"Computer Science and Mathematics Division, Oak Ridge National Laboratory, Oak Ridge, TN 37831}

\title[Models of the quad lens DES J0408-5354]{Models of the strongly lensed quasar DES J0408-5354} 
\author[Agnello et al.]{
\noindent A.~Agnello$^{1}$\thanks{\aaemail},
  H.~Lin$^{2}$,
  L.~Buckley-Geer$^{2}$,
  T.~Treu$^{3\dag}$, V.~Bonvin$^{4}$, F.~Courbin$^{4}$,\and
   C.~Lemon$^{5},$ T.~Morishita$^{3},$
  A.~Amara$^{6},$ M.W.~Auger$^{5},$ S.~Birrer$^{6,3},$ J.~Chan$^{7,8},$\and T.~Collett$^{9},$ A.~More$^{10},$
  C.D.~Fassnacht$^{11},$ J.~Frieman$^{2},$ P.J.~Marshall$^{12},$ R.G.~McMahon$^{5,13},$\and G.~Meylan$^{4},$ S.H.~Suyu$^{8},$ F.Castander$^{14},$ D.~Finley$^{2},$ A.Howell$^{15},$C.~Kochanek$^{16},$\and
  M.~Makler$^{17},$ P.~Martini$^{16},$ N.~Morgan$^{18},$  B.Nord$^{2},$ F.~Ostrovski$^{6,19},$ P.~Schechter$^{20},$\and
  D.~Tucker$^{2},$ R.~Wechsler$^{21},$ T. M. C.~Abbott$^{22},$ F.~B.~Abdalla$^{23,24}$
  S.~Allam$^{2},$\and A.~Benoit-L{\'e}vy$^{25,26,23},$ E.~Bertin$^{25,26},$ D.~Brooks$^{23},$ D.~L.~Burke$^{12,27},$ A. Carnero Rosell$^{28,29},$\and M.~Carrasco~Kind$^{30,31},$ J.~Carretero$^{32},$ M.~Crocce$^{33},$ C.~E.~Cunha$^{12},$ C.~B.~D'Andrea$^{34},$\and L.~N.~da Costa$^{28,29},$ S.~Desai$^{35},$ J.~P.~Dietrich$^{36,37},$ T.~F.~Eifler$^{38},$ B.~Flaugher$^{2},$\and P.~Fosalba$^{14},$ J.~Garc{\'i}a-Bellido$^{39},$ E.~Gaztanaga$^{14},$ D.~A.~Goldstein$^{40,41},$ D.~Gruen$^{12,27},$ \and R.~A.~Gruendl$^{{30,31}},$ 
  J.~Gschwend$^{28,29},$ G.~Gutierrez$^{2},$ K.~Honscheid$^{16,42},$ D.~J.~James$^{22,43},$\and
   K.~Kuehn$^{44},$ N.~Kuropatkin$^{2},$ T.~S.~Li$^{2,45}$ M.~Lima$^{28,46},$ M.~A.~G.~Maia$^{28,29},$ M.~March$^{34}$\and J.~L.~Marshall$^{45},$ P.~Melchior$^{47},$ F.~Menanteau$^{30,31},$ R.~Miquel$^{48,49},$ R.~L.~C.~Ogando$^{28,29}$\and A.~A.~Plazas$^{38},$ A.~K.~Romer$^{50},$ E.~Sanchez$^{51},$ ,R.~Schindler$^{27}$ M.~Schubnell$^{52},$\and I.~Sevilla-Noarbe$^{51},$ M.~Smith$^{53},$ R.~C.~Smith$^{22},$ F.~Sobreira$^{28,54},$ E.~Suchyta$^{55},$\and M.~E.~C.~Swanson$^{31},$ G.~Tarle$^{52},$ D.~Thomas$^{9},$ A.~R.~Walker$^{22}$
  \medskip\\
  $^1$\eso\\
  $^2$\fnal\\
  $^\dag$ Packard Fellow.  The full list of affiliations can be found at the end of the paper.
}

\begin{document}

\voffset-1.in

\date{Accepted . Received }

\pagerange{\pageref{firstpage}--\pageref{lastpage}} 

\maketitle

\label{firstpage}

\begin{abstract}
We present gravitational lens models of the multiply imaged quasar DES
\Jxy, recently discovered in the Dark Energy Survey (DES) footprint,
with the aim of interpreting its remarkable quad-like configuration.
We first model the DES single-epoch $grizY$ images as a superposition
of a lens galaxy and four point-like objects, obtaining  spectral energy distributions (SEDs)
and relative positions for the objects. Three of the point sources (A,B,D)
have SEDs compatible with the discovery
quasar spectra, while the faintest point-like image (G2/C) shows significant reddening and a `grey' dimming of $\approx0.8$mag.
In order to understand the lens configuration, we fit different models to the relative positions of
A,B,D. Models with just a single deflector predict a
fourth image at the location of G2/C but considerably brighter and
bluer. The addition of a small satellite galaxy ($R_{\rm E}\approx0.2^{\ase}$) in the lens plane
near the position of G2/C suppresses the flux of the fourth image and can explain both the reddening and grey dimming.
  All models predict  a main deflector with Einstein radius between $1.7''$ and $2.0'',$ velocity dispersion
$267-280$km/s and enclosed mass $\approx 6\times10^{11}M_{\odot},$
even though higher resolution imaging data are needed to break
residual degeneracies in model parameters. The longest time-delay (B-A) is
estimated as $\approx 85$  (resp. $\approx125$) days by models with (resp. without) a perturber near G2/C.
 The configuration and predicted time-delays of \Jxy~ make it an excellent target for follow-up aimed at understanding the source quasar host galaxy and substructure in the lens, and measuring cosmological parameters. We also discuss some lessons learnt from \Jxy~ on lensed quasar finding strategies, due to its chromaticity and morphology.

\end{abstract}

\section{Introduction}

\begin{figure*}
 \centering
 \includegraphics[width=0.99\textwidth]{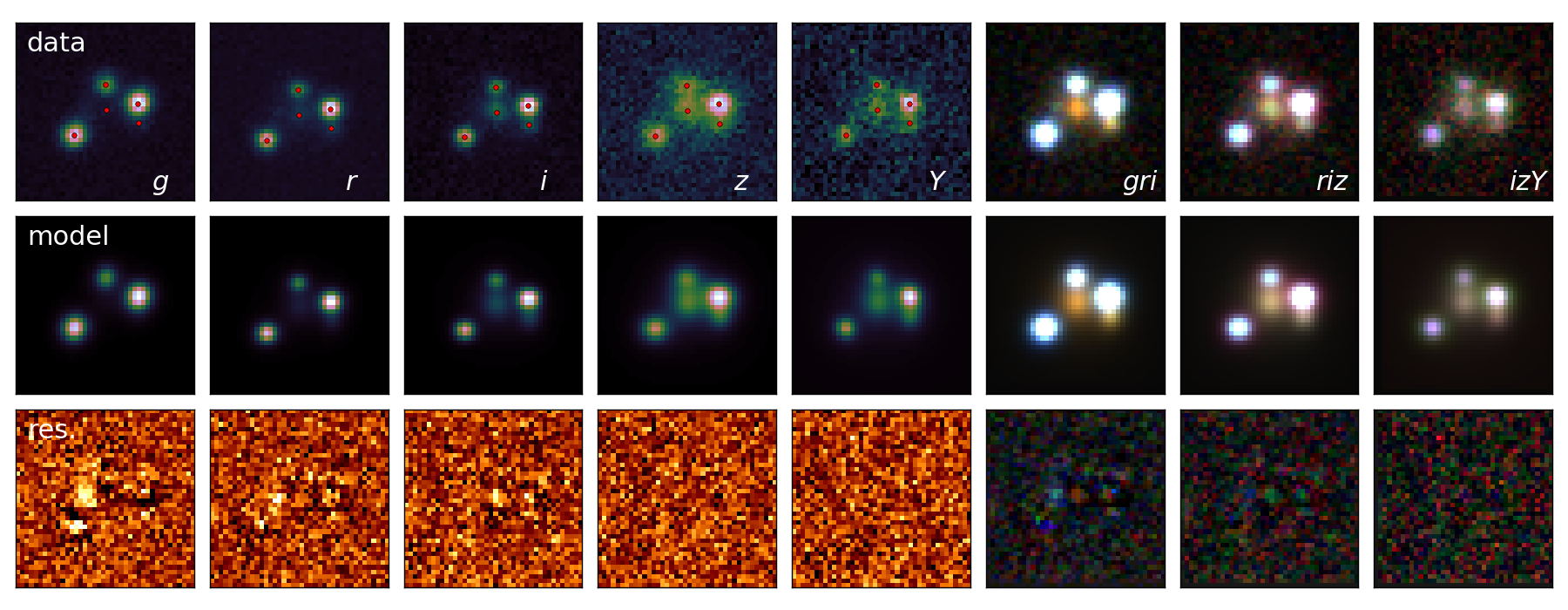}
\caption{\small{Multi-band images of \Jxy~ in $grizY,$ from DES single-epoch data with best image quality, plus colour-composites ($gri,$ $riz,$ $izY$) in the last three columns. The data are shown in the first line with overlaid best-fit positions, the best-fit model (as detailed in Sect.~2) in the second line, and the residuals in the third line. An extra source between A and D is visible in the residuals, indicated as `G3' in fig~2. Most of the residuals, besides G3, are due to PSF mismatch (around image A) and by blending of B and G2/C. North is up and East is left.
 }}
\label{fig:puppies}
\end{figure*}

Strongly lensed quasars are interesting astrophysical objects for diverse purposes \citep{cou02}. The morphology of the multiple images, accompanied by arcs or rings tracing the lensed host galaxy, enables the description of the mass profile of the lens galaxy, which typically sits at redshifts $z_{l}\approx 0.5-1$ \citep[e.g.][]{ogu14}. Thanks to magnification, the source  can be super-resolved, well beyond what is possible for unlensed distant quasars. Astrometric and flux-ratio `anomalies' among the multiple images are signatures of luminous and/or dark substructure surrounding the lens \citep{dal02,nie14}, as well as faint features such as extended disks or isophotal twist, boxiness or diskiness \citep{mol03,mor09,veg12,hsu16,gil16}. When the source luminosity varies over time, the time delay between different images can be measured \citep[e.g.][]{sch97,tew13,bon16} and used to measure cosmological distances \citep[as originally envisioned by][for lensed Supernovae]{ref64} and hence the expansion rate of the Universe, yielding low-redshift ($z_{l}$) constraints on cosmological parameters that are independent of local distance-scale calibrations \citep[cf.][and references therein]{TM16,suy16}.

Image-configuration has a central role for these studies. Systems with four images of the source quasar (hereafter \textit{quads}) provide more information on the mass profiles of the deflector. In contrast, systems with two well-separated images (or \textit{doubles}) can generally be more easily monitored for time variability with ground-based long-cadence observations, since fewer point sources must be de-blended within the same region. Systems in a \textit{fold} configuration, where two of the quasar images are close to one another, are an interesting transition case, that allows for both robust time-delay measurements and lens mass reconstruction \citep{din16}. In particular, in a fold configuration the source lies close to the {caustic} separating the {double} and {quad} regimes, with a \textit{merging pair} of two of the images, thereby giving a highly stretched view of the quasar host near its centre \citep{mor09,rus14,agn16}.

Wide-field surveys offer a significant opportunity to discover new systems with suitable configuration, to be followed up for ancillary data. In particular, the Dark Energy Survey \citep[hereafter DES:][]{san10} has opened a new window for lens searches in the Southern Hemisphere, thanks to a combination of large footprint, depth and good image quality of the Dark Energy Camera \citep{fla15,des16}.

Here, we detail the first models of a new quasar lens, \Jxy~ ($\mathrm{RA}$=62.091333, $\mathrm{DEC}$=-53.900266). This lens was discovered by \citet{lin16} in the Y1A1 release of DES \citep{des14,drl17}, through a visual inspection of blue objects near red galaxies. Its multi-band images show four compact sources, compatible with being point-like given the DES point-spread-function (PSF), around a luminous red galaxy as shown in Figure~1. A spectroscopic confirmation campaign \citep{lin16} shows that the three bright, blue point sources are images of the same source quasar at redshift $\zs=2.375,$ with absorption features at $\zl=0.597$ that can be attributed to the lens galaxy. The fourth compact source to the South-West (fig.~1) is redder than the other confirmed quasar images. Detailed modeling is required to determine whether the anomalous colour is given by dust extinction, microlensing, or an additional red galaxy along the line of sight.

In this follow-up paper, we aim to shed light on the lensing nature of \Jxy, expanding upon the discovery paper. First, we model the DES images to obtain object positions and spectral-energy distributions (SEDs). The multi-band SEDs of the point-sources can be used to quantify chromatic effects (such as microlensing or dust extinction), while the SED of the lens galaxy is used to estimate its stellar mass. The image positions are used as inputs to gravitational lens models, whose results are then used to estimate the dark matter content of the lens and verify whether an additional galaxy, lying very close to the reddened compact source along the line of sight, is needed to reproduce the observed flux ratios. We will show that based on the data available so far, the most plausible interpretation of the system consists of a main deflector galaxy and a satellite producing four images of a background lens quasar. The satellite deflector is very well aligned with one of the images, suppressing its flux and contaminating its colours.

This paper is structured as follows. In Section~2 we detail the multi-band model results of the DES $grizY$ images. A comparison of different lens models is given in Section~3. We conclude in Section~4, including a discussion of the significance of \Jxy~ for different quasar lens searches, and briefly summarize in Section~5. Whenever needed, a standard flat $\Lambda$CDM cosmology is adopted with $\Omega_{\Lambda}=0.7$ and $H_{0}=70\rm{km/s/Mpc}.$
\section{System Configuration}
\begin{figure}
 \centering
\includegraphics[width=0.45\textwidth]{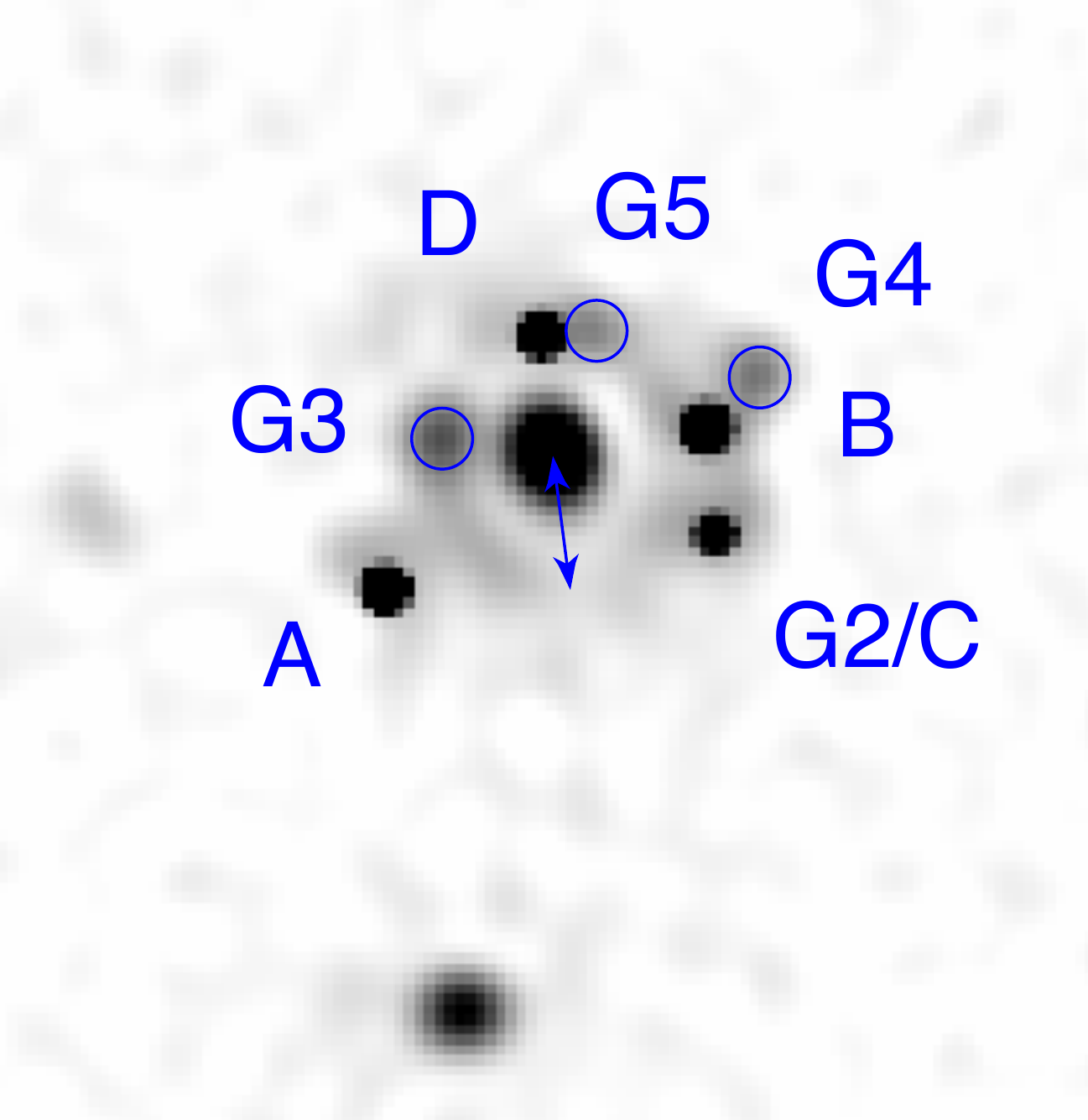}
\caption{\small{WFI $R_{c}-$band image of \Jxy~ after coadd and deconvolution of the best-seeing images.
The central lens galaxy is $G1.$ Image $A,$ farther from $G1,$ is the first to reach the observer, followed by image $B,$ C and $D;$ lens models will need a perturber G2 near the location of image C (see Sct.~3); three additional blobs are visible, marked by blue circles and denoted as G3,G4,G5. G3 sits on a nearly complete ring of radius $\approx1.6^{\ase},$ indicated by the blue arrow.
 }}
\label{fig:WFI}
\end{figure}
\begin{table*} 
\centering
\begin{tabular}{|c||c|c||c|c|c|c|c|}
\hline
obj. & $\delta$RA($^{\ase}$) & $\delta$DEC($^{\ase}$) & $g$ & $r$ & $i$ & $z$ & $Y$\\
\hline
A & 0.0 & 0.0 & 20.07$\pm$0.07 & 20.16$\pm$0.07 & 20.16$\pm$0.07 & 19.96$\pm$0.10 & 20.04$\pm$0.10\\
B & -6.34 & 1.85 & 19.98$\pm$0.07 & 19.95$\pm$0.07 & 19.74$\pm$0.10 & 19.28$\pm$0.08 & 19.34$\pm$0.10\\
G2/C & -6.43 & 0.75 & 22.68$\pm$0.20 & 21.98$\pm$0.15 & 21.46$\pm$0.15 & 20.91$\pm$0.12 & 20.56$\pm$0.16\\
D & -3.12 & 2.91 & 20.90$\pm$0.07 & 20.94$\pm$0.10 & 20.73$\pm$0.12 & 20.42$\pm$0.10 & 20.77$\pm$0.13\\
G1 & -3.31 & 1.48 & 22.18$\pm$0.20 & 20.65$\pm$0.03 & 19.77$\pm$0.04 & 19.31$\pm$0.03 & 19.12$\pm$0.05\\
\hline
A & 0.00 & 0.00 & 20.08$\pm$0.01 & 20.15$\pm$0.01 & 20.15$\pm$0.02 & 19.90$\pm$0.07 & 19.95$\pm$0.14\\
B & -6.35 & 1.86 & 19.86$\pm$0.01 & 19.79$\pm$0.01 & 19.66$\pm$0.02 & 19.29$\pm$0.07 & 19.25$\pm$0.15\\
G2/C & -6.42 & 0.69 & 23.16$\pm$0.11 & 21.61$\pm$0.05 & 20.92$\pm$0.06 & 20.82$\pm$0.09 & 20.45$\pm$0.10\\
D & -3.13 & 2.96 & 20.86$\pm$0.02 & 20.98$\pm$0.02 & 20.90$\pm$0.03 & 20.34$\pm$0.07 & 20.51$\pm$0.15\\
G1 & -3.31 & 1.58 & 22.61$\pm$0.16 & 20.52$\pm$0.06 & 19.51$\pm$0.06 & 19.34$\pm$0.07 & 19.12$\pm$0.08\\
G3 & -1.10 & 1.63 & 22.09$\pm$0.16 & 21.80$\pm$0.17 & 21.50$\pm$0.21 & $>$21.20 & $>$20.85\\
\hline
\end{tabular}
\caption{Positions (relative to image $A$) and SEDs of the objects in \Jxy, from a joint model of the DES $grizY$ single-epoch images with best image quality, adopting the DES-reconstructed PSF (\textit{upper} sub-table) or a parametric fit to a nearby star (\textit{lower} sub-table). Image $A$ is at $(\rm{RA},\rm{DEC})=(62.091323, -53.900289).$ All the positions have an uncertainty of $0.25\times10^{-4}\rm{deg}=0.09^\ase,$ smaller than half the DES pixel size ($0.27^\ase$), with zero covariance between $\delta\rm{RA}$ and $\delta\rm{DEC}.$ The naming scheme is illustrated in Figure~\ref{fig:WFI}. With the current depth and image quality, there are degeneracies in the fitted parameters of G3 and those of other components, primarily G1. The $zY$ magnitudes of the `blue plume' G3 are quoted as upper limits.}
\label{tab:SEDs}
\end{table*}
\Jxy~ consists of point-like and extended objects (fig.~\ref{fig:puppies}), which are blended in the DES segmentation maps. In order to obtain robust SED measurements, in this Section we forward-model the $grizY$ image cutouts as a superposition of objects, to recover robust magnitudes and relative positions with realistic uncertainties.

Follow-up imaging observations are being conducted with the Wide-Field Imager (WFI) on the 2.2m telescope in La Silla, to measure the time-delays between the light-curves of different images. A coadd and optimal deconvolution \citep[following][]{mag98} of the best-seeing images obtained so far, shown in Figure~\ref{fig:WFI}, reveals a more complex structure: besides G1, A, B, G2/C and D, at least three additional `blobs' are visible (G3,G4,G5), as well as a nearly complete Einstein ring with radius $\approx1.6^{\ase}.$ Better data are needed to ascertain the nature of this ring and whether G3, G4, G5 are physically connected to it. The $R_{c}-$band image in Figure~\ref{fig:WFI} has pixels of $0.12^{\ase}$ per side and point-sources with a FWHM$=0.2^{\ase},$ allowing to locate the position angle (p.a.) of G1 to $\approx30$deg E of N. We will discuss these aspects further in the following Sections.

\subsection{Image models}
As illustrated in Figures~\ref{fig:puppies} and \ref{fig:WFI}, the system consists of a red galaxy (G1) surrounded by three blue point-like objects (A, B, D) and a redder and compact object (G2/C). As will be quantified in Section~3, if the system is a genuine quad, then G2/C would be a saddle-point image `C', merging with B in a fold-like configuration. Given the ordering of stationary points in the Fermat potential of a fold configuration \citep{sah03}, the shortest arrival time corresponds to image A (minimum), followed by B (minimum) and C (first saddle-point) and then D (second saddle-point).  For this reason, `C' will be alternatively denoted as the \textit{first saddle-point image} hereafter. Throughout this paper, we will treat this fourth image as an independent object, i.e. will not use its properties directly in constraining the lens models.

The DES cutouts are modelled as the superposition of a galaxy with a \citet{ser68} profile for G1, and four point sources for A,B,D and G2/C. Different choices for the PSF are available, as it can be adopted from the DES PSF reconstruction or explicitly modelled as a superposition of analytic profiles.
Each of these leads to a slight PSF mismatch on pixel-scales, but does not change the results appreciably. In order to test the robustness of the results, we opted for: (\textsc{i}) a model with the DES-reconstructed PSF; and (\textsc{ii}) a model with a Moffat profile \citep{mof69} fit to a nearby star to determine a parametric PSF.
In the model, we impose that the relative displacements of all components (with respect to image A) are the same in every band. The model then comprises: the position angle $\phi_{l},$ S{\'e}rsic index $n_{s}$ and half-light radius $R_{\rm eff}$ of G1; the $grizY$ positions of A; the relative displacements of G1, B, G2/C and D; and the $grizY$ magnitudes of all objects. The Moffat PSF model (\textsc{ii}) includes G3.

\begin{figure}
 \centering
 \includegraphics[width=0.4\textwidth]{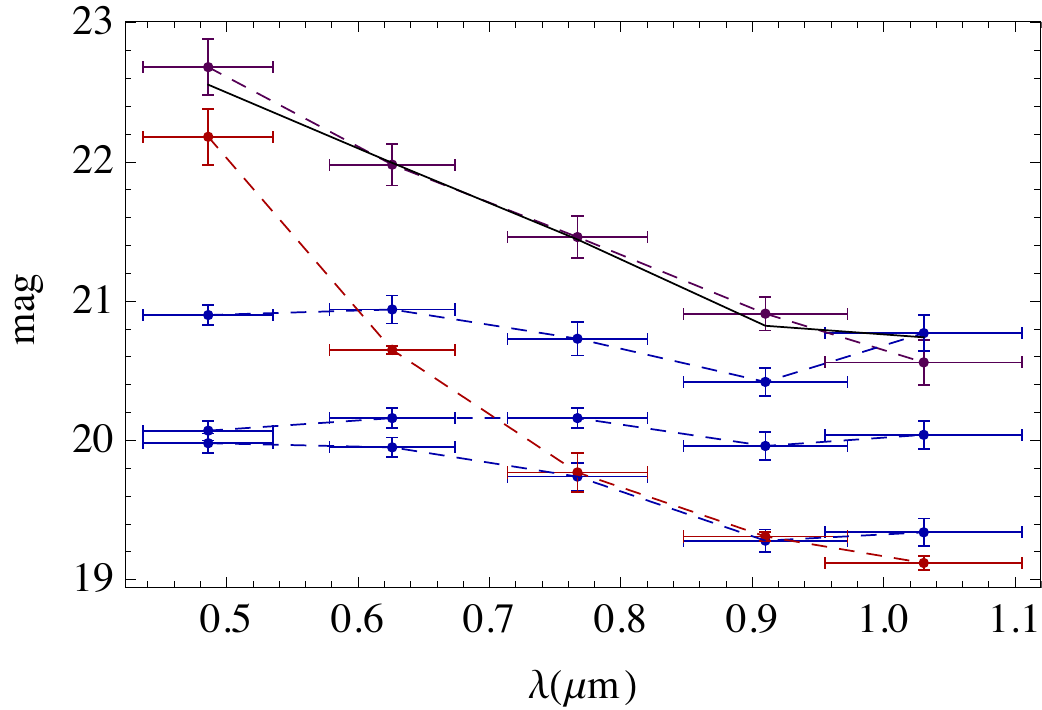}\\
 \includegraphics[width=0.45\textwidth]{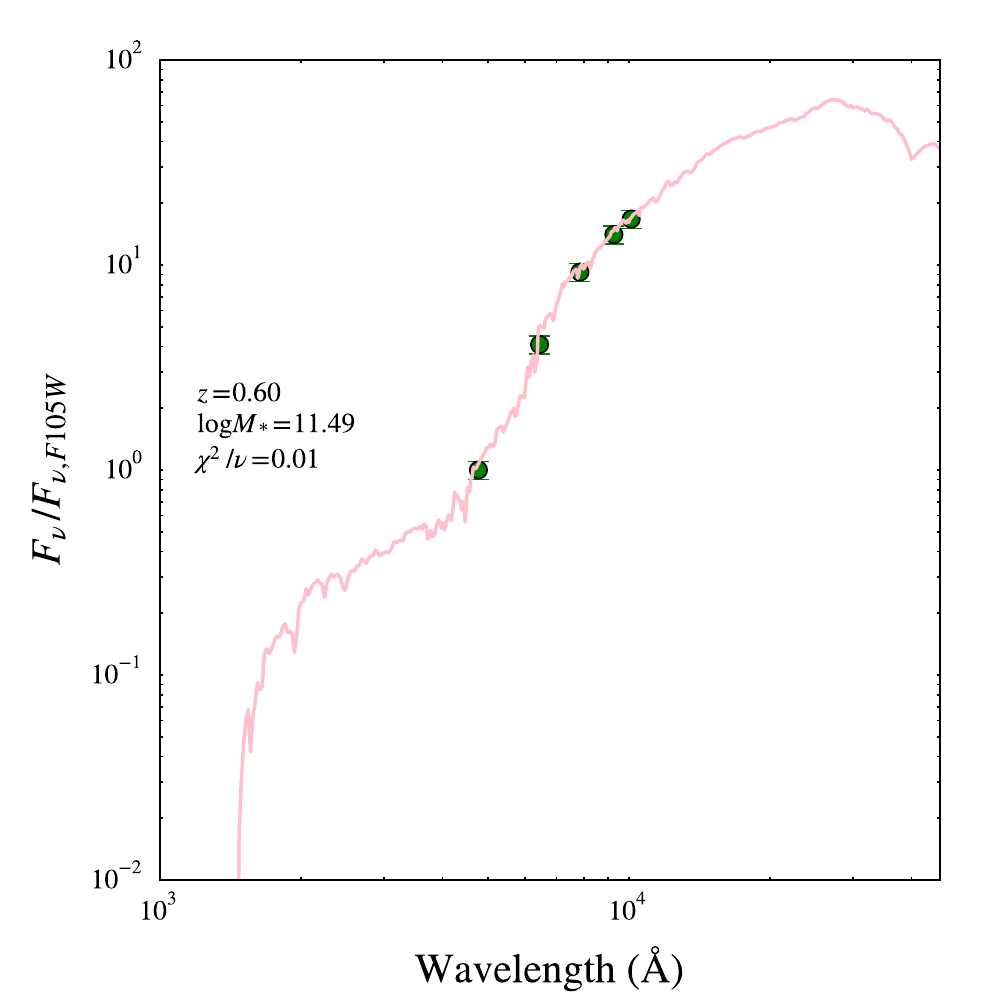}
\caption{\small{\textit{Top:} $grizY$ magnitudes of the multiple components; red (resp. blue) symbols indicate the galaxy G1 (resp. other compact images A,B,D), while the fainter SED with purple symbols corresponds to G2/C. The SED of image B, once reddened, needs an additional dimming of $\approx0.8$mag in all bands to coincide with that of G2/C (black line), to which we also sum the contribution of a galaxy 3.5mas fainter than G1 as discussed in Section 4. \textit{Bottom:} Spectrum of the main deflector galaxy G1 from the best-fit \textsc{FAST} model, yielding $\log_{10}(M_{\star}/M_{\odot})=11.49^{+0.46}_{-0.32}.$ The observed photometry is given by the dark-green symbols. }}
\label{fig:taka}
\end{figure}

The inferred parameters with their uncertainties are listed in Table~\ref{tab:SEDs}. Unfortunately, the depth and image quality of the survey cutouts are not sufficient to constrain $n_{s}$ and $R_{\rm{eff}}.$ Nevertheless, the multi-band magnitudes of G1 are still well constrained. The (broad-band) SEDs of G1 and the four images are shown in Figure~\ref{fig:taka}. The colours of image G2/C can be obtained by adding a standard reddening law \citep[][using $R_V=3.1$ and $E(B-V)=0.3$]{cmm89} to the SED of image B, but the overall magnitudes need an additional `grey' dimming of $0.8$mags; we also sum the small contribution of a putative galaxy G2 3.5mags fainter than G1, in order to better reproduce the $zY-$band fluxes. We will return to these points in Section~3.

\subsection{Lens stellar mass}

The $grizY$ SED inferred for the main deflector galaxy G1 can be used to estimate its stellar mass. We used the public\footnote{Available at \texttt{http://w.astro.berkeley.edu/{\textasciitilde}mariska/FAST.html}} version of \textsc{FAST} \citep{kri09}. Motivated by \citet{tre10b}, we adopt a Salpeter stellar IMF, which is expected for massive early-type galaxies. A direct measurement of the lens velocity dispersion would enable an IMF-independent determination of the stellar mass \citep{aug09}. The best-fit model is shown in Figure~\ref{fig:taka}. With the uncertainties from the SED modelling, we obtain $\log_{10}(M_{\star}/M_{\odot})=11.49^{+0.46}_{-0.32}.$ We will compare this to the results of lens models in the next Section.

\section{Lens models}

\begin{table*} 
\centering
\begin{tabular}{|c||c|c|c|c|c|c|c||c|}
\hline
 & $\theta_{{\rm E}, l}$ & $q$ & $\phi_l$ & $\gamma_{s}$ & $\varphi_{s}$ & $b_{p}$ & $s_{p}/b_{p}$ & $\theta_{{\rm E},p}$\\
\hline
SIE & $(1.98\pm0.08)^\ase$ & $0.63\pm0.06$ & $-60.0\pm2.0$ & --- & --- & --- & --- & --- \\
SIS+XS & $(1.87\pm0.08)^\ase$ & [1.00] & --- & $0.13\pm0.3$ & $29.8\pm3.3$ & -- & --- & --- \\
SIS+XS+pert. & $(1.73\pm0.15)^\ase$ & [1.00] & --- & 0.11$\pm$0.03 & 18.4$\pm$10.1 & $(0.33\pm0.23)^\ase$ & 0.24$\pm$0.21 & $(0.26\pm0.13)^\ase$\\
\hline
SIS+XS+pert.$^{(a)}$ & $(1.72\pm0.10)^\ase$ & [1.00] & --- & 0.10$\pm$0.02 & 16.5$\pm$7.2 & $(0.35\pm0.19)^\ase$ & 0.30$\pm$0.20 & $(0.22\pm0.08)^\ase$\\
\hline
\end{tabular}
\caption{Inferred lens model parameters in the case of a Singular Isothermal Ellipsoid (SIE, first row), a Singular Isothermal Sphere plus external shear (SIS+XS, second row), or the same plus a small perturber near $G2,$ adopting $0.2^\ase$ for the positional uncertainties of $A,B,D$ and $0.3^\ase$ for those of G1 and G2. The lens p.a. of G1 (which may be different from that of its starlight) is quoted in `mathematical notation' N of W, corresponding to $\approx 30$ deg E of N. The perturber Einstein radius $\theta_{{\rm E},p}$ is not an additional parameter, being inferred directly from $b_{p}$ and $s.$ Models with a sub-critical G2 ($s>2b_{p}$) are not excluded. $^{(a)}$The last line shows the average parameters and  standard deviations obtained when all uncertainties on positions are set to $0.1^\ase.$
}
\label{tab:lenspars}
\end{table*}
\begin{table*} 
\centering
\begin{tabular}{|c||c|c|c|c||c|c|}
\hline
  model & $\log_{10}\mu(A)$ & $\log_{10}\mu(B)$ & $\log_{10}\mu(D)$ & $\log_{10}\mu(C)$ & $x_{\rm{C}}-x_{\rm{G1}}$ ($^\ase$) & $y_{\rm{C}}-y_{\rm{G1}}$ ($^\ase$)\\
\hline
SIE & 0.47 & 0.89 & 0.52 & 0.76 & 1.65$\pm$0.05 & -0.89$\pm$0.03 \\
     & $0.45\pm0.09$ & $0.93\pm0.17$ & $0.45\pm0.18$ & $0.86\pm0.20$ &  &   \\
SIS+XS & 0.52 & 1.05 & 0.58 & 1.14 &1.60$\pm$0.05 & -0.70$\pm$0.05  \\
            & $0.64\pm0.13$ & $1.09\pm0.15$ & $0.71\pm0.17$ & $0.93\pm0.19$ &  &   \\
\hline
SIS+XS+pert. & 0.64 &  0.82 & 0.51 & 0.49 & 1.58$\pm$0.03  & -0.47$\pm$0.07  \\
                     & $0.77\pm0.19$ &  $0.97\pm0.14$ & $0.62\pm0.16$ & $0.70\pm0.25$ &  &  \\
\hline
\end{tabular}
\caption{Inferred \textit{logarithmic} magnifications for the three models, with one SIE (top) or SIS+XS (middle) in the lens plane, or with the addition of a perturber (bottom). The first line of each block is $\log_{10}(\mu)$ from the best-fit model, while the second line shows the mean and standard deviation from the MCMC posterior. The last column lists the predicted displacement of image $C,$ in terms of West-ward and North-ward displacements from the best-fitting position of $G1$ from Sect.~2 (identified with $\delta x= \delta y =0$). The positional uncertainties are systematics-dominated, as the predicted position (especially $y_{\rm C}$) can change appreciably across models.
}
\label{tab:mags}
\end{table*}

The three images A, B, D have compatible SEDs, as is also confirmed by their long-slit spectra by \citet{lin16}. We can then use their positions relative to G1 to model this system as a gravitational lens, obtaining estimates of the total mass (within the Einstein radius) and predicted time-delays between different images. Since G2 is substantially redder than the other components, we do not include it in the lens model, but rather compare its properties with those predicted by the lens model fit to the other components. The technicalities of the lens model are described in Appendix A.

Conservatively, we adopt $0.2^\ase$ positional uncertainties on $A,B,D$ and $0.3^\ase$ on G1, G2, about twice as large as those from the cutout modelling of Section~2 (relying solely on the DES cutouts). This allows us to explore a wide family of lens models and draw some general conclusions, in particular on the flux-ratios allowed by different models. In one case, we also allow the positional uncertainties to be those given directly by the cutout modelling (last line of tab.~\ref{tab:lenspars}). The inferred lens model parameters for all models are given in Table~\ref{tab:lenspars}. We stress that we are not using the smaller uncertainties from the WFI deconvolution, in order to highlight the robustness of some conclusions that held already with DES-quality data. However, when ellipticity is included in the lens model (defined as `SIE' below), its p.a. agrees well with that from the WFI images shown in Figure~\ref{fig:WFI}.

The images A, B, D are mapped to the source plane according to the lens equation
\begin{equation}
\boldsymbol{\theta}_{s}=\boldsymbol{\theta}_{im}-\boldsymbol{\alpha}-\boldsymbol{\Gamma}\boldsymbol{\theta}_{im}\ ,
\end{equation}
where $\boldsymbol{\theta}=(\delta x, \delta y)$ is the angular displacement relative to the best-fitting G1 center from Section~2, the \textit{external shear} matrix $\boldsymbol{\Gamma}$ is defined as
\begin{equation}
\boldsymbol{\Gamma} = \gamma_{s}\, \left( \begin{array}{cc}
\cos(2\varphi_{s}) & \sin(2\varphi_{s}) \\
\sin(2\varphi_{s}) & -\cos(2\varphi_{s}) \end{array} \right)\ 
\end{equation}
and $\boldsymbol{\alpha}$ depends on how we describe the deflections by lensing galaxies.
When describing lens galaxies, we use parametric models for their convergence profiles $\kappa=\Sigma/\Sigma_{cr},$ where $\Sigma_{cr}={\rm c}^{2}D_{s}/(4\pi {\rm G}D_{l}D_{ls})$ accounts for the dimensional dependence on angular-diameter distances.
In particular, we use a Pseudo-Isothermal Ellipsoidal Mass Profile \citep[PIEMD,][]{kas93}. This model provides a good representation of the gravitational potential of lens galaxies (e.g. Treu 2010) and the deflection angles $\boldsymbol{\alpha}$ in coordinates $(X,Y)$ aligned with the principal axes of the iso-density ellipsoids
 \begin{eqnarray}
 \alpha_{X}\ =\ -\frac{b}{\sqrt{1-q^{2}}} \arctan\left(\frac{X\sqrt{1-q^{2}}}{s+\sqrt{q^{2}(s^{2}+X^{2})+Y^{2}}}\right)\\
 \alpha_{Y}\ =\ -\frac{b}{\sqrt{1-q^{2}}} {\rm arctanh} \left(\frac{Y\sqrt{1-q^{2}}}{q^{2}s+\sqrt{q^{2}(s^{2}+X^{2})+Y^{2}}}\right)
 \end{eqnarray}
are fully analytic, together with the convergence and the Fermat potential. The expression in coordinates $(x, y)$ in West-North orientation requires just rotations in the coordinates and deflections, for which we choose the lens long-axis p.a. $\phi_l$ as positive N of W. The spherical ($q\rightarrow1$) and core-less ($s/b=0$) limit reduces to the Singular Isothermal Sphere (SIS), for which $b$ is also the Einstein radius $R_{\rm E}$ enclosing a mean convergence of 1. In the Singular Isothermal Ellipsoid (SIE) case ($q<1,$ $s/b=0$), with the above notation we have $R_{\rm E}=b/\sqrt{q}$ as the ellipsoidal coordinate of the contour enclosing $\langle \kappa \rangle=1.$ In the case where $q=1$ but $s/b>0,$ the Einstein radius is $R_{\rm E}=b\sqrt{1-2s/R_{\rm E}},$ which means that the PIEMD can be sub-critical ($\kappa<1$ everywhere) when $s>b/2.$ The Einstein radius can be used to estimate the lens velocity dispersion via\footnote{The numerical prefactors in the second equalities are specific to the redshifts $z_s,$ $z_l$ of source and deflector in this particular case.}
\begin{equation}
\sigma_{\rm sis}={\rm c}\sqrt{\frac{R_{\rm E}D_{s}}{4\pi D_{l}D_{ls}}}
\ =203\,(\theta_{\rm E}/1^{\prime\prime})^{1/2}\rm{km/s}
\end{equation}
while the projected mass within $R_{E}$ is
\begin{equation}
M_{p}(R_{E})=\pi\Sigma_{\rm cr}R_{E}^{2}
\  = 2.0*10^{11}(\theta_{\rm E}/1^{\prime\prime})^{2}M_{\odot}\ ,
\end{equation}
regardless of the lens model. Here and in what follows, $\theta_{E}=R_{E}/D_{l}$ is the Einstein radius in angular units, the same as for the lens strength parameter $b$.

\subsection{Models with one Deflector}

\begin{figure}
 \centering
 \includegraphics[width=0.22\textwidth]{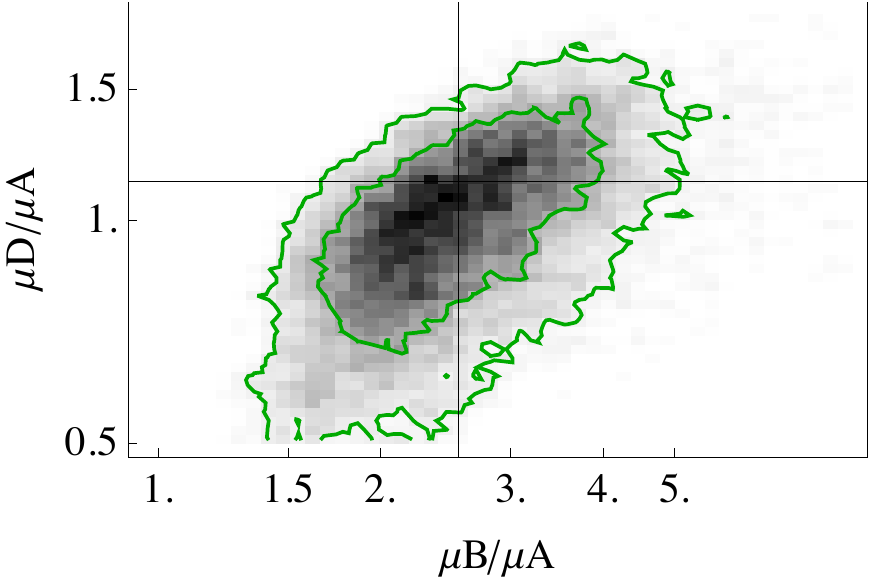}
 \includegraphics[width=0.22\textwidth]{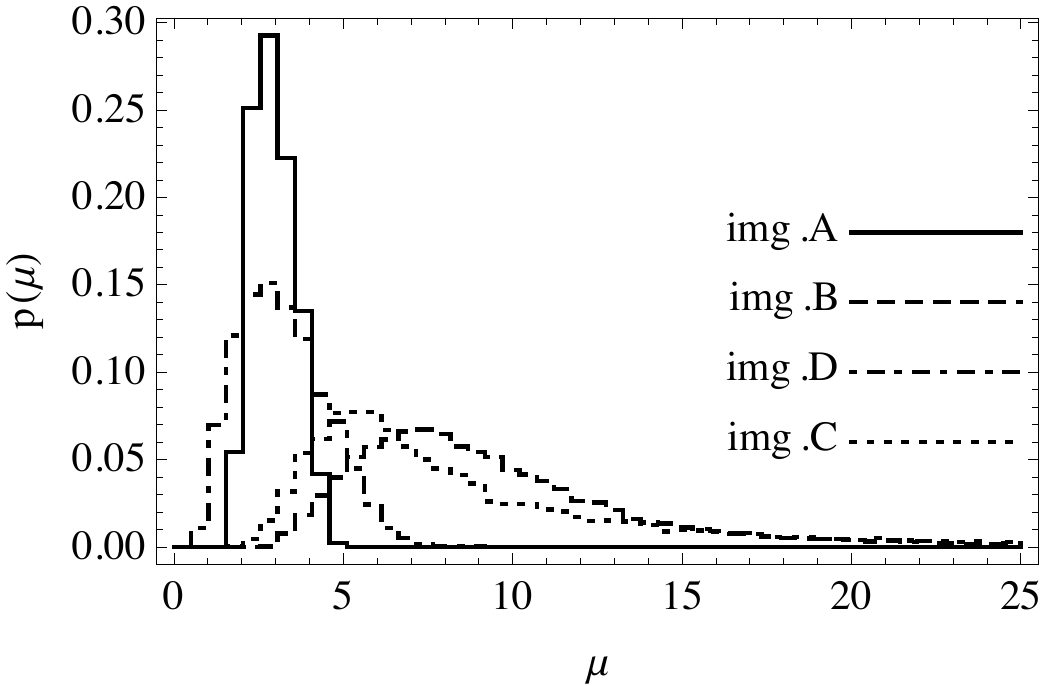}\\
 \includegraphics[width=0.22\textwidth]{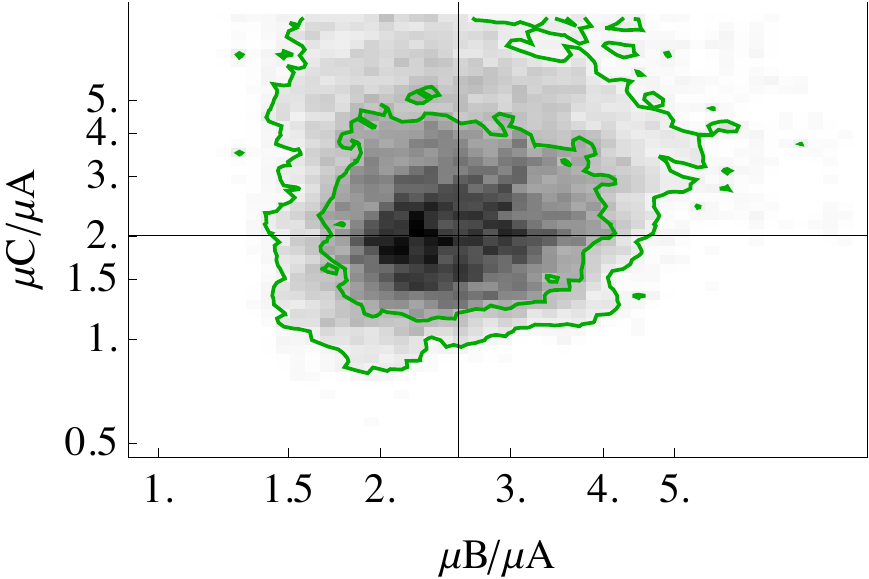}
 \includegraphics[width=0.22\textwidth]{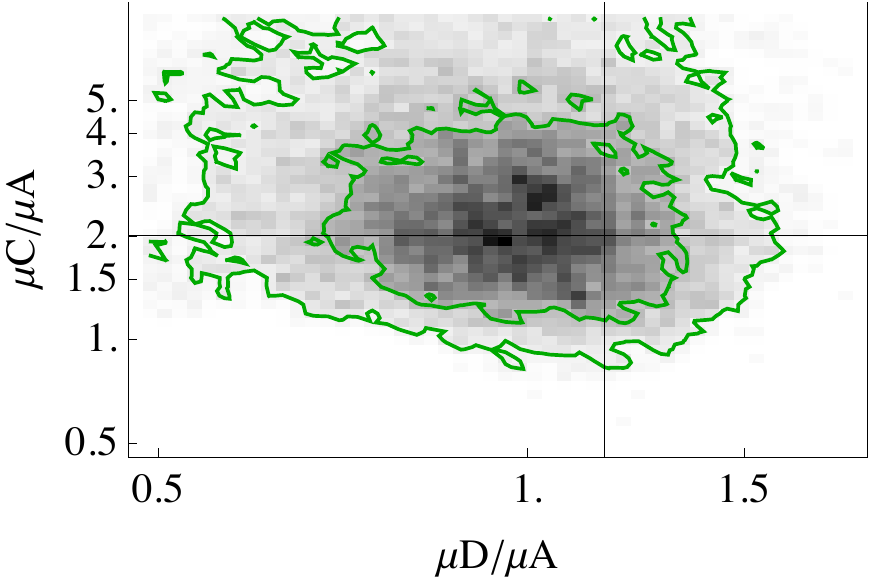}
\caption{\small{
 Output magnifications from a model with one SIE in the lens plane; the green contours represent the $68\%$ and $95\%$ quantiles of the marginalized posterior (no parameters held fixed). While the magnifications of $B$ and $D$ relative to $A$ are in qualitative agreement with the SED fit results, the predicted image $C$ should be almost as bright as image $B$ and appreciably brighter than image $A.$
 This is not observed even after differential reddening is added to fit the colours (Sect.~2), and so it cannot be solely the result of dust extinction.
 }}
\label{fig:outmagsSIE}
\end{figure}

For the first models, we describe the lensing mass distribution as given solely by G1. The first model (SIE) comprises simply a SIE representing G1. The second model (SIS+XS) adopts a SIS for G1, with the addition of external shear with non-null $\gamma_{s}.$ The resulting parameters are listed in Table~2.

\begin{figure*}
 \centering
 \includegraphics[width=0.45\textwidth]{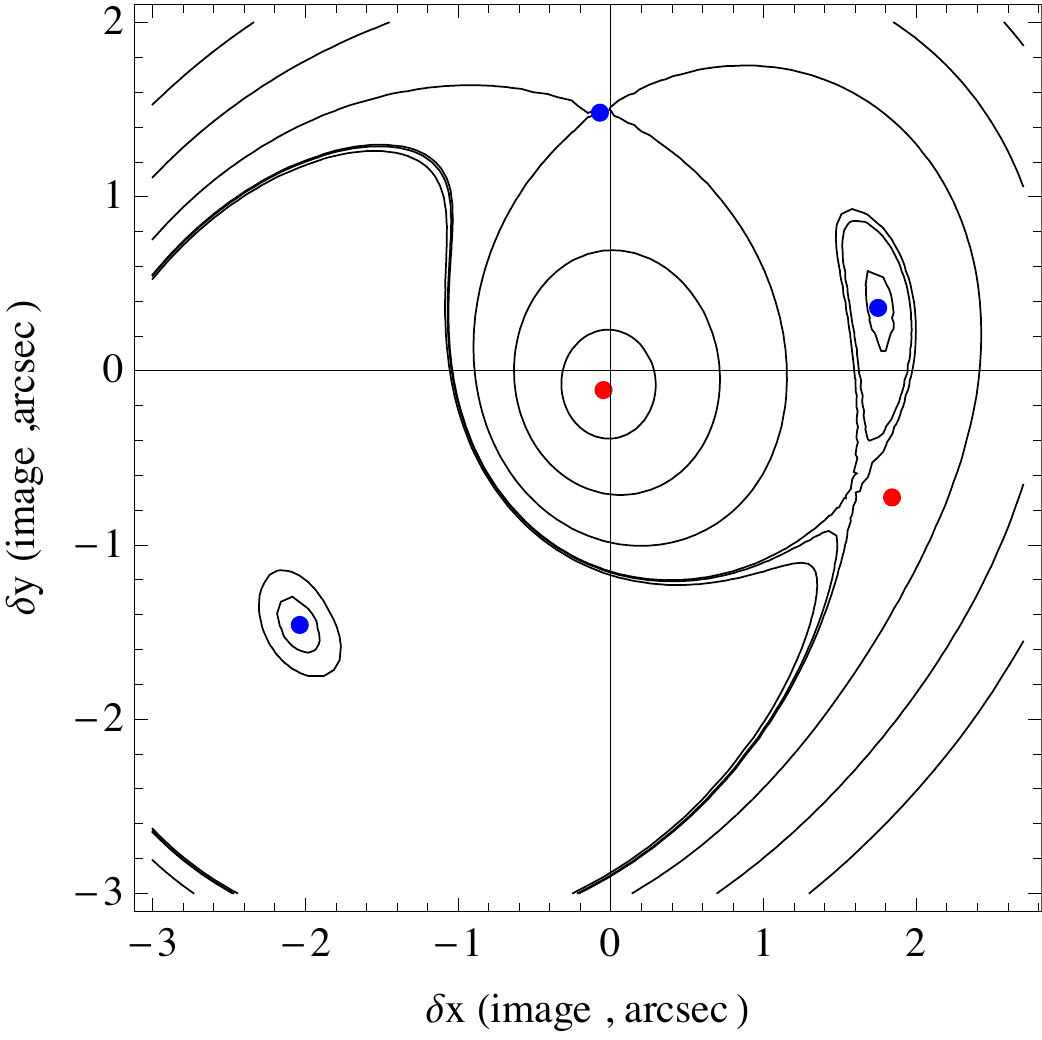}
 \includegraphics[width=0.45\textwidth]{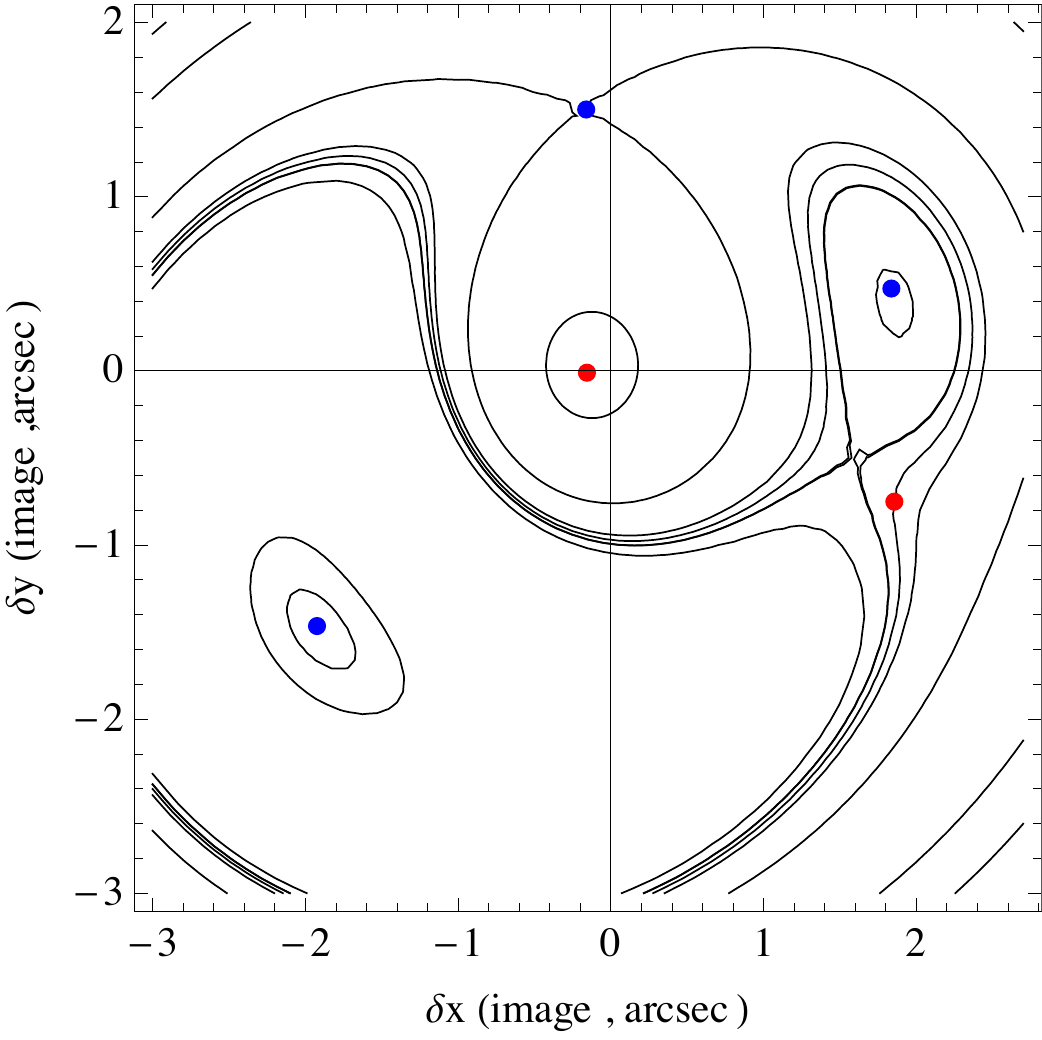}\\
\caption{\small{
 Time-delay contours for the case with one SIS plus external shear (\textit{left} panel) or with the addition of a perturber at $G2$ (\textit{right}). Models generally predict the fourth image position within one pixel-length in each direction from $G2.$ This has a magnification comparable to that of image $B$ if no perturber is present nearby. Being a saddle-point image, its magnification is easily suppressed by the presence of a small perturber at $G2.$
 }}
\label{fig:imageplane}
\end{figure*}

Both the SIE and SIS+XS models reproduce the positions of images A,B,D and predict a saddle-point image `C' near the position G2/C found in Sect.~1 (fig.~\ref{fig:imageplane}), whose relative position can vary from model to model, still within one or two DES pixels. The inferred Einstein radius $\theta_{{\rm E},l}$ of G1 is slightly less than half the A-to-B image separation ($\approx 2.2^\ase$), due to quadrupole contributions to the deflection either by ellipticity or by shear. The quadrupole shear-ellipticity degeneracy is evident in that the shear angle $\varphi_{s}$ in the SIS+XS case is orthogonal to the inferred lens position angle $\phi_{l}$ of the SIE case. The lens velocity dispersion and mass within $R_{\rm E}$ can be estimated as $(286\pm6)$km/s and $(7.9\pm0.6)10^{11}M_{\odot}$ (resp. $(280\pm6)$km/s and $(7.0\pm0.6)10^{11}M_{\odot}$) for the SIE (resp. SIS+XS) model.

\begin{figure}
 \centering
 \includegraphics[width=0.45\textwidth]{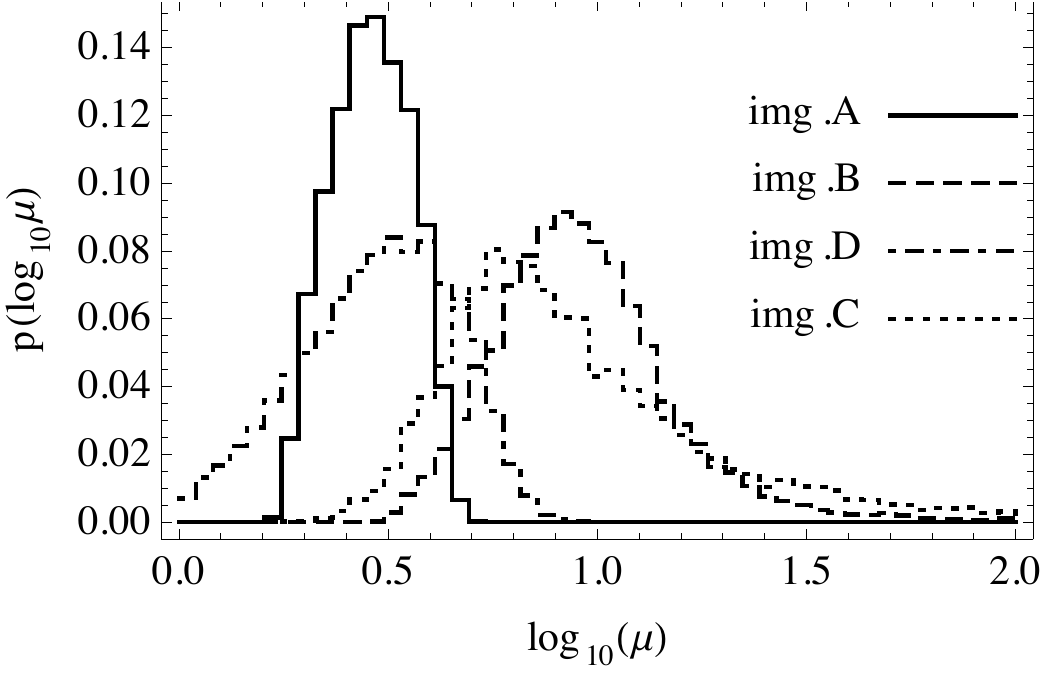}\\
 \includegraphics[width=0.45\textwidth]{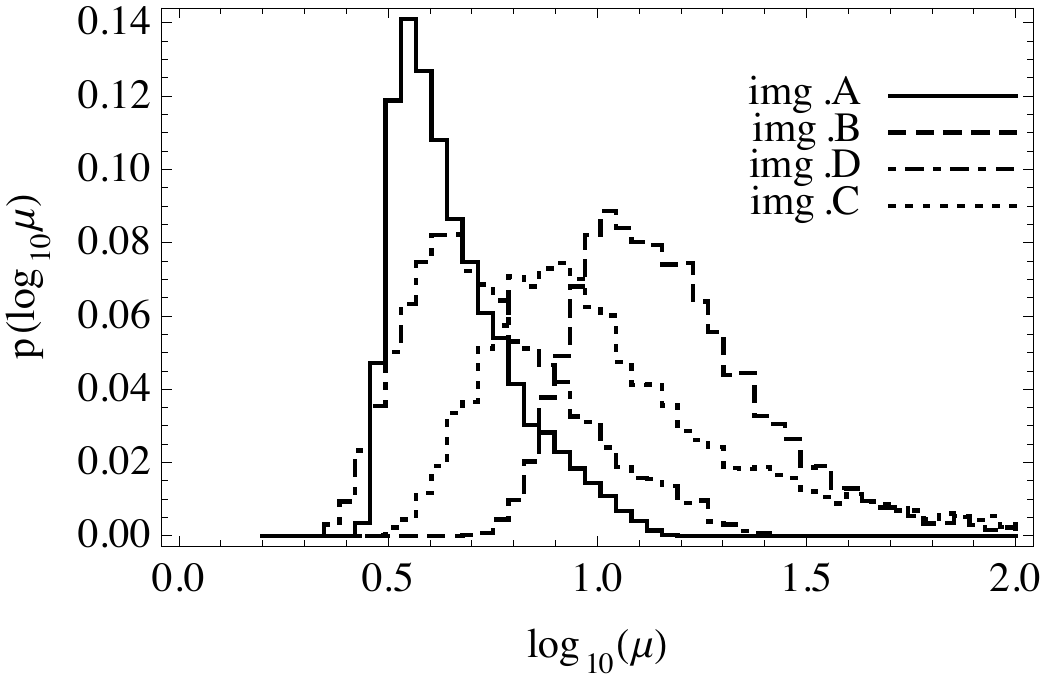}\\
\caption{\small{
    Output magnifications from a model SIE (top) and SIS+XS (bottom), in logarithmic units.
    The SIE and SIS+XS models produce similar results, particularly for the predicted ordering of magnifications.
 }}
\label{fig:outmagswopert}
\end{figure}

Models with just one central deflector predict that image `C' should be about as bright as image `B', even with relatively large {adopted} uncertainties on the image positions (0.2$^\ase$ instead of $0.09^\ase$). This is summarized in Figures \ref{fig:outmagsSIE} and \ref{fig:outmagswopert}, and in Table \ref{tab:mags}.

\subsection{Models with a Perturber}
The first saddle-point predicted by models with one deflector would fall near the position of G2, which however is appreciably redder than the other images and significantly fainter than predicted even in the reddest bands.  Extinction as measured in other lensed quasars \citep{dai06,med05} does not differ substantially to that measured in the Milky Way and Magellanic Clouds \citep[e.g.][]{cmm89}. However, while the simple addition of a standard reddening law\footnote{With $R_V=3.1$ and $E(B-V)=0.3,$ blueshifting the DES wavebands to the lens rest-frame.} to the SED of image B can reproduce the colours of image G2/C, it still requires a `grey' dimming of $\approx0.8$mag in each band to match its overall magnitudes as in fig~\ref{fig:outmagsSIE}.
 
Since G2/C lies close to image B, the differential reddening should be produced by a local overdensity, such as a small galaxy, whose lensing effect can also alter the magnification of image C. In general, saddle-points of the Fermat potential are \textit{suppressed}, i.e. dimmed, by the presence of nearby perturbers, whereas minima fluctuate less \citep{sch02,kee03}. 

For this reason, we add a galaxy at the location of G2/C, which we describe as a PIEMD with $q=1.$ The addition of a perturber at a fixed position increases the number of parameters by two (core size and Einstein radius), making the model under-constrained. However, we can still rely on the priors on positions given by Section~2, and examine the range of parameter configurations that are compatible with the observed image configuration.

For simplicity, and due to the lack of an independent redshift measurement, we place the perturber in the same plane of the main lens G1. In general, models of lenses with four images have degeneracies among the monopole and quadrupole parameters \citep{koc04}. As verified above, the SIS+XS and SIE models do not differ appreciably in the output image positions and magnifications (tab.~\ref{tab:lenspars}, \ref{tab:mags}, fig.~\ref{fig:outmagswopert}).

\begin{figure}
 \centering
 \includegraphics[width=0.22\textwidth]{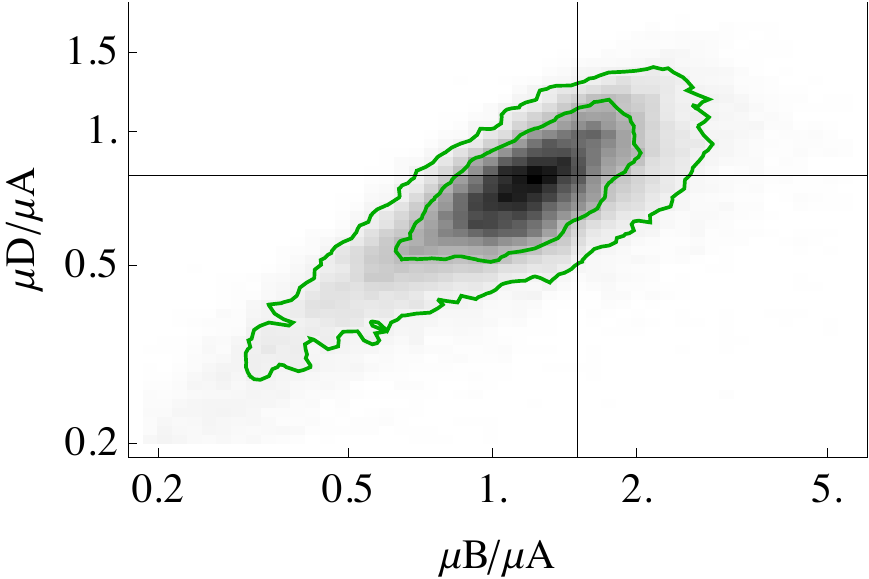}
 \includegraphics[width=0.22\textwidth]{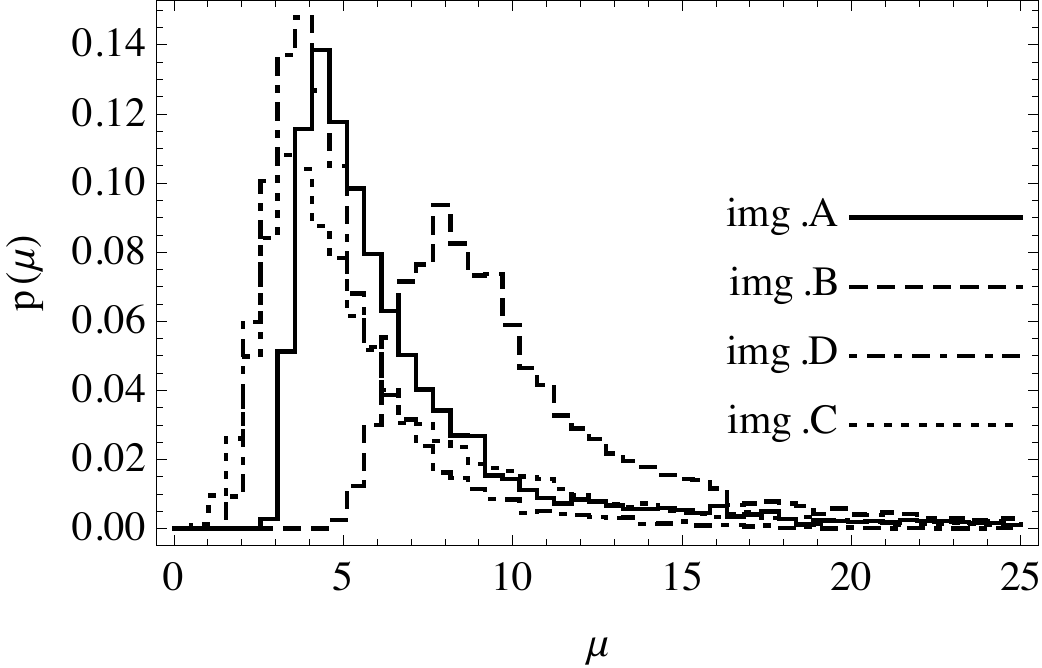}\\
 \includegraphics[width=0.22\textwidth]{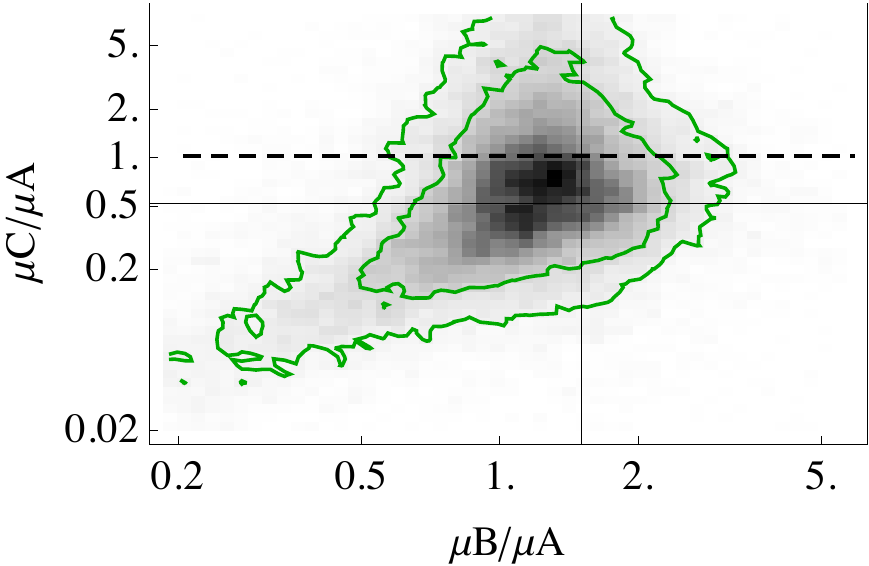}
 \includegraphics[width=0.22\textwidth]{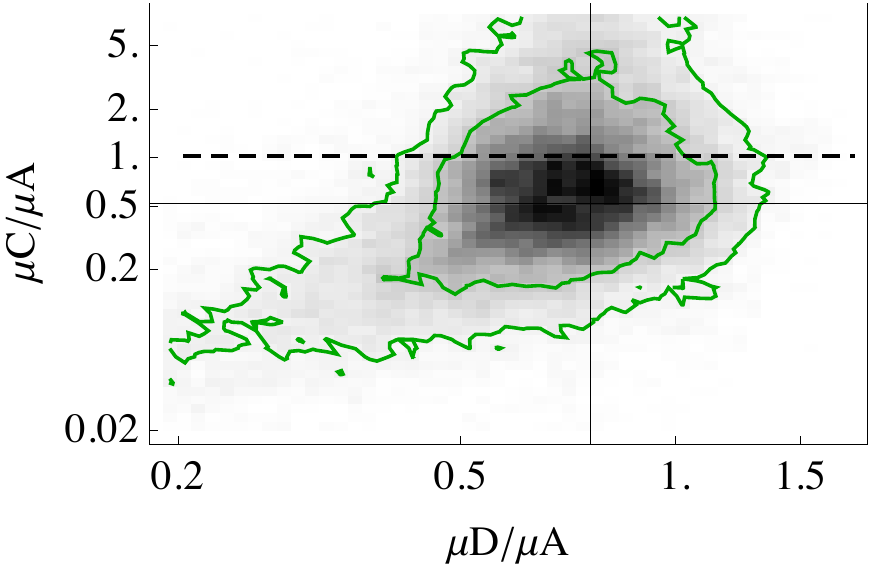}\\
\caption{\small{
 Output magnifications from a model with SIS+XS plus a small perturber near position $G2.$; again, the green contours represent the $68\%$ and $95\%$ quantiles of the marginalized posterior.
 This model predicts relative magnifications that are in agreement with the flux ratios obtained from the SEDs obtained in Section~2, with both images $C$ and $D$ slightly fainter than image $A$ and significantly fainter than image $B$.
 }}
\label{fig:outmagswpert}
\end{figure}

The inferred lens parameters of the new model (with a main lens G1 and a perturber G2), given in Table~\ref{tab:lenspars}, suggest a fairly small ($\approx 0.2^\ase$) Einstein radius and do not rule out a sub-critical perturber. Similarly to the findings of \citet{nie14} on a different lens, these limits are given simply by the requirement that the other images (A,B,D) are not shifted by the perturber beyond their measured uncertainties.

With the addition of G2 in the lens model, the output magnifications are in agreement with what is measured in Sect.~1, and the predicted image `C' (fig.~\ref{fig:imageplane}) is suppressed by the presence of the small perturber, making it slightly fainter than image A (fig.~\ref{fig:outmagswpert}). With a small perturber at $\approx 0.2^{\ase}$ from image C East-ward and North-ward, its SED can be easily reddened even though it lies very close to B. The small separation between C and G2 makes them hardly distinguishable even in the Gemini acquisition image of \citet{lin16}, whose PSF has a quoted FWHM$\approx 0.5^{\ase}.$ Within this model, the lens velocity dispersion of G1 is $(267\pm12)$km/s, and its projected mass within $R_{\rm E}$ is $(6.0\pm1.0)10^{11}M_{\odot}.$ Even though G2 is not excluded to be sub-critical, we can still estimate its velocity dispersion and enclosed projected mass as $(95\pm17)$km/s  and $\lesssim 1.0\times10^{10}M_{\odot},$ respectively.

\begin{figure}
 \centering
 \includegraphics[width=0.22\textwidth]{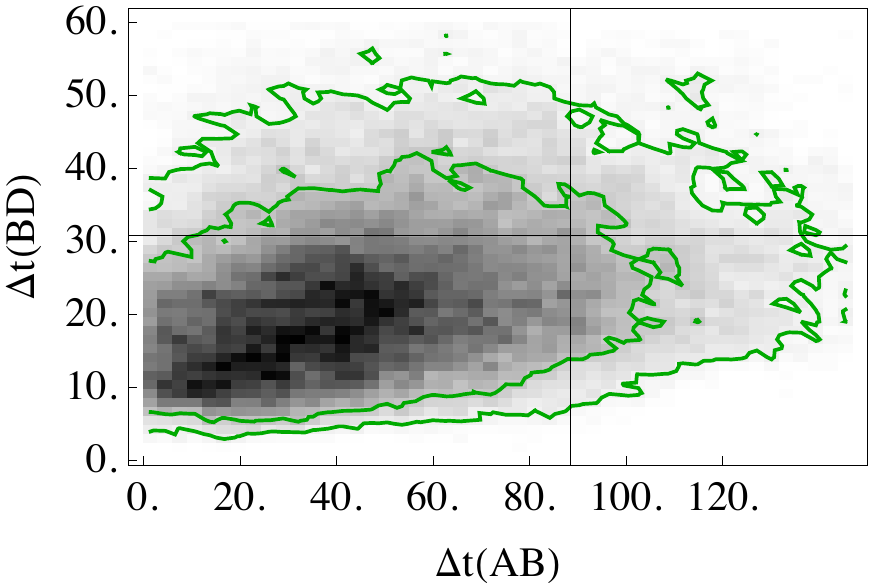}
 \includegraphics[width=0.22\textwidth]{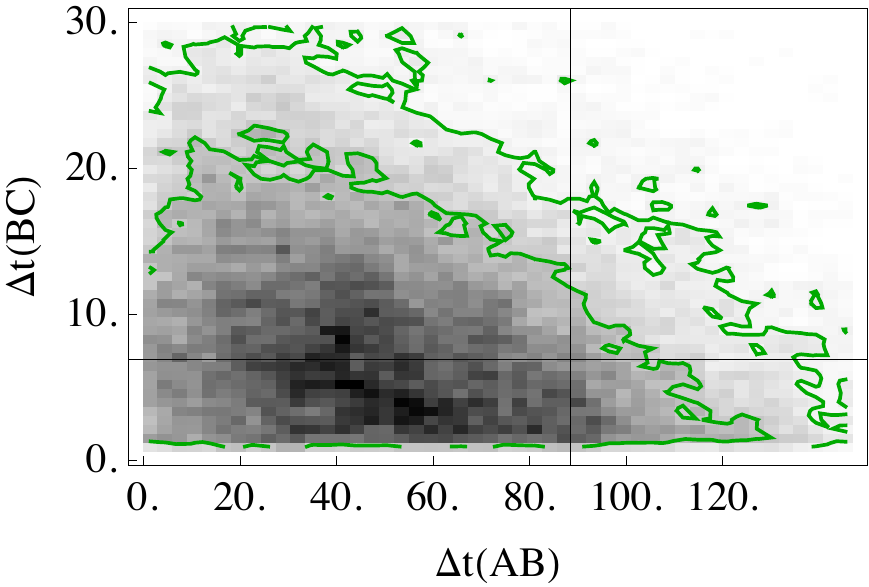}\\
\caption{\small{
    Predicted time delays between different quasar images. Most of the inferred values are offset from the results of the best-fitting lens model, which predicts $\Delta t(AB)=85$d, $\Delta t(CB)=6$d, $\Delta t(BD)=29$d.
 }}
\label{fig:tdels}
\end{figure}

For this model, we also give some forecasts on the expected time delays from the arrival times
\begin{equation}
t_{i}=\frac{(1+z_{l})D_{l}D_{s}}{\mathrm{c}D_{ls}}\left[\frac{1}{2}\left|\boldsymbol{\theta}_{im,i}-\boldsymbol{\theta}_{s}\right|^{2}-\Phi\right]\ ,
\end{equation}
where the projected potential $\Phi$ is analytic in all models chosen. Within the SIS+XS+perturber model, we have
$\Delta t(AB)=85$d, $\Delta t(BC)=6$d, and $\Delta t(BD)=29$d, where $\Delta t(i,j)=t_{j}-t_{i}$ is positive when the arrival-time of image $i$ is shorter than that of image $j$. The quoted values have large uncertainties, due to the wide degeneracies in the lens models, and their marginalized posterior is offset from the values from the best-fitting model (fig.~\ref{fig:tdels}). The ordering of time-delays is general and does not depend on whether a perturber is included in the model, being  determined by the configuration of critical points \citep[e.g.][]{sah03}.
Delays like those of \Jxy~ are ideal for ground based monitoring, because they are long enough to yield 1-2\% precision with daily cadence, yet short enough that one or two observing seasons are sufficient.

\section{Discussion}
%
We have modelled \Jxy~ to obtain the photometry of its individual components, the stellar mass of the main lens galaxy and lens parameters for a choice of plausible models. The predicted time-delays and image configuration make this system amenable to follow-up for time-delay cosmography, as well as for studies of the quasar host near the central engine and substructure near the quasar images.

With the current data quality, there are vast degeneracies in the lens model parameters, which however can be easily relieved with high-resolution imaging data. This will also help locate the pertuber G2 responsible for the reddening \textit{and} dimming of image C. The occurrence of both cases would not be uncommon, as seen e.g. for the lens B1608+656 \citep{mye95,fas96,suy09}. A direct measurement of the lens velocity dispersion, together with a follow-up campaign for time-delays, would yield a direct measurement of the angular-diameter distance to the lens via $D_{l}\propto \Delta t/\sigma^{2}$ \citep[see][for a general discussion]{par09,jee16}. 

\subsection{System Configuration}
\Jxy~ consists of three blue point-like images of the source quasar at $z_s=2.375,$ and two redder components of which G1, the main lens, is at $z_l=0.597,$ whereas the nature of G2/C is unclear, given its colours and the significant contamination from spectra of other components (Lin et al. 2016). We have modelled this system as a superposition of an extended galaxy (G1) plus four compact sources A, B, D, G2/C and obtained deconvolved SEDs. In particular, the SED of G1 suggests a stellar mass $M_{\star}\approx 3.2\times10^{11}M_{\odot}$ for the lens, within $\approx0.4$dex uncertainty. There is a degree of systematic uncertainty in the positions and fluxes of B and G2/C, given their proximity and the fact that B is more than a magnitude brighter than G2. Some faint residuals are given by PSF mismatch on pixel scales. The WFI images with best seeing, once deconvolved, show what could be an additional source that is mapped into a nearly complete ring with $R_{\rm E}\approx1.6^{\ase},$ which is slightly but appreciably smaller than that inferred from lens models based on images A,B,D (as summarized below).

\subsection{Lens Model Properties}
The positions of images A, B, D relative to G1 have been used to explore lens models of \Jxy. Models with one main lens, adopted as Singular Isothermal Ellipsoid or Singular Isothermal Sphere plus external shear, predict an Einstein radius $\approx1.9^\ase$ and a lens p.a.$\approx-60$deg North of West, or $0.1$ shear orthogonal to it. Both models, while successful at reproducing the positions of images A, B, D, would predict a saddle-point image where G2 lies and about as bright as image B, which is not observed even in band Y. Models with a small perturber at the location of G2 reproduce the same image positions, but are able to suppress image C by about a magnitude. The magnitude difference $2.5\log_{10}(\mu_{C}/\mu_{B})\approx0.8$ is in very good agreement with the grey dimming found in Section~2.

The projected mass within the Einstein radius is $M_{p}(R_{\rm E})\approx 6\times 10^{11}M_{\odot},$ about twice the stellar mass estimated from the SED of G1. A proper evaluation of the dark matter fraction in the lens, however, would require a measurement of the efective radius of G1. When the perturber has non-null Einstein radius, its enclosed mass is $M_{p}(G2)\approx 1.3\times10^{10}M_{\star}.$
 The contribution of a small galaxy with magnitudes $m(G2)=m(G1)-2.5\log_{10}(M_{p}(G2)/M_{p}(G1))$ is barely noticeable in $gri$ bands, which in turn can be well reproduced by reddening and offsetting the SED of image B, and makes the $zY-$band magnitudes of image G2/C in complete agreement with the values measured from Section 2 (black line in fig.~\ref{fig:taka}).

The estimated time-delay between images A and B is $\approx85$ days, making this system amenable to follow-up for time-delay cosmography. Still, given the uncertainties on image positions and few constraints, the derived uncertainties are sizeable and higher-resolution imaging data will be required to tighten the model-predicted uncertainties on the delays.

If indeed two sources are present at different redshift, \Jxy~ can also be used to measure Dark Energy cosmological parameters via the ratio of distance ratios $D_{s}/D_{ls}$ to the different sources \citep[e.g.][]{pac81,suc04,col12}, besides time-delay cosmography to measure $H_0$. The only other system with time delays and multiple source-planes that is known and studied to date is the galaxy cluster MACSJ1149.5+2223 \citep[][]{tre16}.

\subsection{Relevance of \Jxy~ for lens searches}
The photometry and configuration of \Jxy~ make it an interesting testbed for different techniques of lensed quasar candidate selection.  These, in turn, have implications for substructure studies, as the composition of lens-selected or source-selected samples affects the sensitivity to substructure, especially for lens searches that are tailored on simple lenses or on systems dominated by `isolated' quasar SEDs.

Like the serendipitous quad of \citet{anu16}, \Jxy~ was originally found by visual inspection of objects selected solely on $gri$ survey properties, instead of relying on hybrid infrared `excess' colours \citep{war00} that have been used to target quasars \citep{mad06, mad12,pet15} or lenses \citep{ofe07} and applied in other lens searches in DES \citep{agn15,ost16}. After the initial discovery via the blue-near-red search of \citet{lin16}, different teams have examined their own search methods. Here we provide a summary of the different findings.

\subsubsection{Cutout classification: CHITAH}
\textsc{CHITAH} \citep{cha15} examines the image cutouts of objects to detect at least two blue compact sources and a red galaxy, evaluating how plausible the configuration is as a strong lens via the corresponding source-plane $\chi^2$. This approach relies on the requirement that the blue images have very similar SEDs, distinct from the lens SED.

When applied to the $grizY$ cutouts of \Jxy, it did not flag this system as a possible quad since the fourth image G2/C is significantly redder than the others. However, based on A, B and G1, it did classify this system as a possible double. These findings suggest that pixel-based automatic recognition, such as \textsc{CHITAH} or \textsc{LensTractor}\footnote{Available at \texttt{https://github.com/davidwhogg/LensTractor}} could be made more flexible by accounting for possible SED variations of the predicted images.

\subsubsection{Target selection: data mining}

The first technique used to select lensed quasars in the DES relied upon Artificial Neural Networks (ANNs) trained on SDSS $griz$ and WISE \citep{wri10} $W1,W2$ bands of four main classes of objects \citep{agn15}. Despite the success of the first discovery results \citep{agn15b}, further improvements could be made for wider application to DES, as discussed in Appendix B.
With these new ANNs, \Jxy~ was automatically flagged as an extended quasar with $z_{s}>1.75,$ one of the two classes (besides `lens') to be retained for visual inspection\footnote{In particular, the blend D+G1 with catalogue ID=3070264166, RA=62.0904688061, DEC=-53.8996413857}. Despite the improvement in the ANNs and the \textit{blind} re-discovery of \Jxy, there is considerable scatter in the SDSS-DES translated magnitudes, which can cause some interesting objects to slip out of the selection boundaries (and false positives to leak in). The outlier selection method (Agnello et al. in prep.), in which \Jxy~ is rediscovered as a $>3\sigma$ outlier among quasars and with low probability to be a galaxy, is somewhat immune from this issue, as are Population Mixture classifications \citep{ost16,wil16}.


\section{Summary}

\Jxy~ has an interesting fold-like image configuration, with three well-separated images (A,B,D) and a fourth one (C) in a merging pair with the brightest image (B).
Besides the three, clearly identifiable blue images of the source quasar, a fourth component G2/C is fainter than simple lens-model predictions and appreciably red. While image B is already redder than the farthest image A, with $\Delta(Y-r)\approx0.65$ compatible with a simple \citep{cmm89} reddening law with $E(B-V)=0.3,$ image G2/C is further reddened (additional $E(B-V)=0.3$) {and} also requires a grey dimming of $0.8$mag in every band. 

A small perturber ($R_{\rm E,p}\approx 0.23^\ase$, $M_{p}\approx1.0\times10^{10}M_{\odot}$) near the location of G2/C explains both the needed reddening and dimming over the whole $grizY$ range.

The image separation makes this system particularly apt to time-delay measurements, with an expected B-A delay $\approx85$days. The lens mass within the Einstein radius $R_{\rm E}=1.73^\ase$ is $M_{p}\approx(6.0\pm1.0)\times10^{11}M_{\odot},$ about twice the stellar mass of the main galaxy G1 $M_{\star}\approx3\times10^{11}M_{\odot}.$

The chromaticity and morphology of \Jxy~ mean that different search techniques, while successfully flagging it as a lens candidate, are triggered by different features. Also, the peculiar colours and configuration of the quasar images are a powerful reminder that automated search techniques should be flexible enough to encompass these systems, in view of homogeneous lens-selected or source-selected samples for follow-up science. \citet{om10} estimated $1146$ quasar lenses within a depth of $i=23.6$ in the 5000deg$^2$ final DES footprint, of which $14\%$ quads. Past and ongoing lens searches show that a suite of complementary techniques are needed to maximize the number of detected lenses, especially at magnitudes fainter than $i\approx 19.$

The composition of \Jxy, with a primary (massive) lens and a small perturber and a merging image-pair, make it both an interesting system for follow-up and a rather peculiar system to model. Spectroscopic and high-resolution imaging observations would enable more accurate models, both for cosmography and for substructure studies, and a highly magnified view of the source quasar and its host.

\section*{Acknowledgments}

AA and TT acknowledge support by the Packard Foundations through a Packard
Research Fellowship and by the National Science Foundation through
grant AST-1450141.

This paper was written as part of the STRong lensing Insights into the Dark Energy Survey (STRIDES) collaboration, a broad external collaboration of the Dark Energy Survey,
 \texttt{http://strides.astro.ucla.edu}
 
%
Funding for the DES Projects has been provided by the DOE and NSF(USA), MISE(Spain), STFC(UK), HEFCE(UK). NCSA(UIUC), KICP(U. Chicago), CCAPP(Ohio State),  MIFPA(Texas A\&M), CNPQ, FAPERJ, FINEP (Brazil), MINECO(Spain), DFG(Germany) and the Collaborating Institutions in the Dark Energy Survey.  The Collaborating Institutions are Argonne Lab, UC Santa Cruz, University of Cambridge, CIEMAT-Madrid, University of Chicago, University College London,  DES-Brazil Consortium, University of Edinburgh, ETH Z{\"u}rich, Fermilab, University of Illinois, ICE (IEEC-CSIC), IFAE Barcelona, Lawrence Berkeley Lab,  LMU M{\"u}nchen and the associated Excellence Cluster Universe, University of Michigan, NOAO, University of Nottingham, Ohio State University, University of  Pennsylvania, University of Portsmouth, SLAC National Lab, Stanford University, University of Sussex, and Texas A\&M University.  The DES Data Management System is supported by the NSF under Grant Number AST-1138766. The DES participants from Spanish institutions are partially  supported by MINECO under grants AYA2012-39559, ESP2013-48274, FPA2013-47986, and Centro de Excelencia Severo Ochoa SEV-2012-0234. Research leading  to these results has received funding from the ERC under the EU's 7$^{\rm th}$ Framework Programme including grants ERC 240672, 291329 and 306478.

\appendix
\section{Lens modeling specifics}
Regardless of the model specifics, all images must map to the same source-position. For each choice of the lens model parameters, a source at $\boldsymbol{\theta}_s$ in the source plane corresponds to images $\boldsymbol{\theta}_i$ in the image plane, and the goodness-of-fit can be described by the image-plane $\chi^2$
\begin{equation}
 \chi^{2}_{ip}=\sum_{i=1}^{3}\frac{\left|\boldsymbol{\theta}_{i}-\boldsymbol{\theta}_{im,i}\right|^{2}}{\delta^{2}_{i}}
= \sum_{i=1}^{3}\frac{\left|\mathbf{A}_{i}(\boldsymbol{\theta}_s-\boldsymbol{\theta}_{s,i})\right|^{2}}{\delta^{2}_{i}}\ ,
\end{equation}
where $\boldsymbol{\theta}_{im,i}$ and $\boldsymbol{\theta}_{s,i}$ are the measured image-positions and their model-predicted source-plane positions for images A,B,D, $\mathbf{A}_{i}=\partial\boldsymbol{\theta}_{im,i}/\partial\boldsymbol{\theta}_{s,i}$ is the magnification tensor around each image and $\delta_{i}$ is the positional uncertainty on image $i.$ The second equality relies on the fact that, near a reasonable lens solution, we can linearize the lens equation around the measured image positions. Its validity has been tested extensively by \citet{ogu10}. Writing the $\chi^{2}$ as above relies on a Gaussian distribution of the measured image positions, with isotropic positional uncertainties, and is equivalent to drawing image positions with infinite precision from Gaussians $\mathcal{G}(\boldsymbol{\theta}_{im,i},\delta_{i}),$
considering (for each choice) a highly-penalized image-plane $\chi^2=p\chi^{2}_{ip,1}$ in the lens model with
\begin{equation}
 \chi^{2}_{ip,1}=\sum_{i=1}^{3}\left|\boldsymbol{\theta}_{i}-\boldsymbol{\theta}_{im,i}\right|^{2}
\end{equation}
and  $p\gg\delta^{-2}_{i}.$ This allows us to generalize the lens model likelihood to image configurations that do not have isotropic and Gaussian uncertainties. In particular, we can draw the relative displacements of G1, B and D with respect to image A as given by the likelihood explored in Section 2, which we call  $\mathcal{L}_{SED}.$ At very high values of $p,$ the only parameter combinations that are explored are those that correspond to all image positions mapping back to the same source position, because other configurations are heavily penalized.

Another hypothesis underlying this approach is that the measured image position uncertainties are simply given by the extraction of Section~2, so that each image carries a weight proportional to its (squared) magnification in the $\chi^2.$ This does not account for systematic uncertainties in the image positions given by the proximity of different objects and PSF mismatch. This problem is evident for the brightest image B, which would instead carry the highest weight in $\chi^{2}_{im}.$
 We then opt for a penalized source-plane $\chi^2$ of the form
\begin{equation}
\chi^{2}_{sp}\ =\ p\sum_{j=1}^{3}\left|\boldsymbol{\theta}_{s,j}-\langle\boldsymbol{\theta}_{s}\rangle \right|^{2}\ ,
\end{equation}
where $\langle\boldsymbol{\theta}_{s}\rangle=(\boldsymbol{\theta}_{s,A}+\boldsymbol{\theta}_{s,B}+\boldsymbol{\theta}_{s,C})/3$ for each choice of the model parameters, and consider the lens-model likelihood as
\begin{equation}
\mathcal{L}\ \propto\ \mathcal{L}_{SED}(\boldsymbol{\theta})\times\rm{e}^{-\chi^{2}_{sp}/2}\ .
\end{equation}
The penalty parameter $p$ is gradually increased, until all possible models are effectively producing images originating from the same source-position, within milli-arcsecond tolerance, and the model uncertainties are driven by $\mathcal{L}_{SED}$.

\section{Mining Across Surveys}
The original implementation of ANNs was based upon SDSS data and four main classes of objects. In order to be more widely applicable to DES, it was improved in three ways: (\textsc{i}) more object classes, including multiple redshift intervals for the `quasar' class to distinguish low-redshift contaminants from higher-redshift objects; (\textsc{ii}) less restrictive colour-cuts, that would otherwise exclude known lenses with higher $g-i$ or lower $W1-W2;$ and (\textsc{iii}) accounting for the differences in photometry between SDSS and DES via a cross-calibration valid for blue extended objects\footnote{We refer to \cite{agn15b} for the definition of `blue and extended' in this case.}. The best-fit regressions have
\begin{eqnarray}
\nonumber g_{\rm des}\ =\ g_{\rm sdss}+0.05,\ \ r_{\rm des}\ =\ r_{\rm sdss}+0.088,\\
i_{\rm des}\ =\ i_{\rm sdss}+0.112,\ \ z_{\rm des}\ =\ z_{\rm sdss}+0.159,
\end{eqnarray}
for the \texttt{psf} magnitudes, and
\begin{eqnarray}
\nonumber g_{\rm des}\ =\ g_{\rm sdss}+0.165-0.092(g_{\rm des}-r_{\rm des}-0.4)\\
\nonumber r_{\rm des}\ =\ r_{\rm sdss}+0.118-0.215(g_{\rm des}-r_{\rm des}-0.4)\\
\nonumber i_{\rm des}\ =\ i_{\rm sdss}+0.04-0.2(i_{\rm des}-z_{\rm des})\\
 z_{\rm des}\ =\ z_{\rm sdss}+0.078-0.044(z_{\rm des}-Y_{\rm des}-0.17)
\end{eqnarray}
for the \texttt{model} magnitudes. There is considerable scatter ($0.11-0.18$mag) in the translated magnitudes, given by the extendedness of the objects and different depth and image quality between SDSS and DES. This means that interesting candidates (resp. contaminants) can leak out of (resp. within) the hyperplanes defining class boundaries as identified by the ANN classification.

\section*{Affiliations}
  $^1$\eso\\
  $^2$\fnal\\
  $^3$\ucla\\
  $^4$\epfl\\
  $^5$\ioa\\
  $^6$\eth\\
  $^7$\asiaa\\
  $^8$\mpa\\
  $^9$\ports\\
  $^{10}$\ipmu\\
  $^{11}$\ucd\\
  $^{12}$\kipac\\
  $^{13}$\kavli\\
  $^{14}$\espa\\
  $^{15}$\ucsb\\
  $^{16}$\ohioa\\
  $^{17}$\cbpf\\
  $^{18}$\nick\\
  $^{19}$\braz\\
  $^{20}$\mit\\
  $^{21}$\stanf\\
  $^{22}$\ctio\\
  $^{23}$\ucl\\
  $^{24}$\rhodes\\
  $^{25}$\cnrs\\
  $^{26}$\sorb\\
  $^{27}$\slac\\
  $^{28}$\braza\\
  $^{29}$\brazb\\
  $^{30}$\illa\\
  $^{31}$\illb\\
  $^{32}$\espb\\
  $^{33}$\espa\\
  $^{34}$\penns\\
  $^{35}$\hyder\\
  $^{36}$\exc\\
  $^{37}$\lmu\\
  $^{38}$\jpl\\
  $^{39}$\espd\\
  $^{40}$\berk\\
  $^{41}$\lbnl\\
  $^{42}$\ohiob\\
  $^{43}$\wash\\
  $^{44}$\aao\\
  $^{45}$\texas\\
  $^{46}$\brazc\\
  $^{47}$\prince\\
  $^{48}$\cata\\
  $^{49}$\ifae\\
  $^{50}$\sussex\\
  $^{51}$\espc\\
  $^{52}$\mich\\
  $^{53}$\south\\
  $^{54}$\abc\\
  $^{55}$\oak\\

\label{lastpage}

\end{document}